\documentclass[a4paper,aps,prd,onecolumn,superscriptaddress]{revtex4-2}
\usepackage{amsmath,amssymb}
\usepackage{graphicx,subfigure,epsfig}
\usepackage{hyperref}
\usepackage{geometry}
\usepackage{tablefootnote}
\usepackage{setspace}

\usepackage{footmisc}

 \usepackage{booktabs}
 \usepackage{multirow}
 \usepackage[normalem]{ulem}
 \useunder{\uline}{\ul}{}

\hypersetup{
    colorlinks=true,
    citecolor=blue,
    linkcolor=red,
    filecolor=magenta,
    urlcolor=cyan,}
\usepackage{color}

\begin{document}

\doublespacing

\title{Echoes of a hairy black hole from gravitational decoupling}

\author{Yi Yang}
\email{yiyang@mail.gufe.edu.cn}
\affiliation{School of Mathematics and Statistics, \\
Guizhou University of Finance and Economics, Guiyang, 550025, China}

\author{Ali \"Ovg\"un}
\email{ali.ovgun@emu.edu.tr}
\affiliation{
Physics Department, Eastern Mediterranean University, Famagusta, 99628 North Cyprus via Mersin 10, Turkey.}

\author{Gaetano Lambiase}
\email{lambiase@sa.infn.it}
\affiliation{Dipartimento di Fisica ``E.R Caianiello'', Universit degli Studi di Salerno, Via Giovanni Paolo II, 132 - 84084 Fisciano (SA), Italy}
\affiliation{Istituto Nazionale di Fisica Nucleare - Gruppo Collegato di Salerno - Sezione di Napoli, Via Giovanni Paolo II, 132 - 84084 Fisciano (SA), Italy}

\author{Dong Liu}
\email{dongliuvv@yeah.net}
\affiliation{Department of Physics, Guizhou Minzu University, Guiyang, 550025, China}

\author{Zheng-Wen Long}
\email{zwlong@gzu.edu.cn}
\affiliation{College of Physics, Guizhou University, Guiyang, 550025, China}

\begin{abstract}
We study axial gravitational perturbations of a hairy black hole constructed within the framework of gravitational decoupling and investigate the geometric origin of echo-like late-time signals in this spacetime. We derive the odd-parity master equation and the corresponding effective potential, and compute the quasinormal mode spectrum using frequency-domain and time-domain methods. We show that the axial potential develops a double-peak structure that supports a trapping cavity and gives rise to echo-like waveforms. The representative double-peak configurations also exhibit an unstable--stable--unstable light-ring structure, which provides the null-geodesic counterpart of the trapping cavity in the eikonal limit. Therefore, the echoes are not produced by an artificially imposed near-horizon condition, but instead emerge dynamically from the geometric structure of the effective potential. Our results provide a useful framework for probing black hole hair through gravitational-wave ringdown and for exploring possible observational departures from the standard no-hair paradigm. 
\end{abstract}

\maketitle
\section{Introduction}

The direct detection of gravitational waves by the LIGO and Virgo detectors has opened an unprecedented observational window onto the strong-field and highly dynamical regime of gravity. In particular, the first binary-black-hole event, GW150914, and the subsequent LIGO--Virgo catalogs have established compact-binary coalescences as precision probes of general relativity in the nonlinear regime, while the post-merger signal has highlighted the physical importance of the ringdown phase as a direct tracer of the geometry of the remnant spacetime \cite{LIGOScientific:2016aoc,LIGOScientific:2017vwq,KAGRA:2023pio}.

Quasinormal modes (QNMs) are the characteristic damped oscillations of perturbed black holes and related compact objects, and they have long served as a central tool in black-hole perturbation theory, ringdown modeling, and gravitational-wave spectroscopy \cite{Berti:2025hly, Nollert:1999ji,Berti:2009kk,Konoplya:2011qq,Buonanno:2006ui,Cardoso:2016rao,Barausse:2014tra,Hod:2006jw,Maggiore:2007nq}. At this stage, the waveform is governed by a discrete set of damped oscillations, known as QNMs, whose complex frequencies encode the mass, angular momentum, and electric charge. This makes the QNM spectrum a central arena for black-hole spectroscopy, no-hair tests, and phenomenological searches for departures from the standard Kerr and Schwarzschild geometries \cite{Dreyer:2003bv,Giesler:2019uxc,Bhagwat:2019dtm}.
Beyond their role in strong-gravity phenomenology, QNMs also exhibit deep connections to holography, angular eigenvalue problems, unstable null geodesics, shadow observables, and semianalytic or fully numerical extraction schemes, as illustrated by studies of holographic modes, spin-weighted spheroidal harmonics, Kerr/eikonal geometry, time-domain evolution, higher-order WKB methods, and continued-fraction techniques \cite{Kovtun:2005ev,Berti:2005gp,Yang:2012he,Dolan:2010wr,Konoplya:2017wot,Andersson:1996cm,Konoplya:2019hlu,Daghigh:2011ty,Daghigh:2024wcl,Li:2026xnr}. A broad body of work has explored QNMs in asymptotically AdS and dS spacetimes, analogue systems, scalar-hairy and regular black holes, charged and nonlinear-electrodynamic backgrounds, phase-transition settings, scale-dependent geometries, and several modified-gravity scenarios, including Einstein--Gauss--Bonnet, Einstein--dilaton--Gauss--Bonnet, slowly rotating vector perturbations, loop-quantum-corrected black holes, and Weyl black holes \cite{Cardoso:2001bb,Wang:2004bv,Daghigh:2008jz,Zhidenko:2003wq,Lepe:2004kv,Zhidenko:2005mv,Fernando:2012yw,Toshmatov:2015wga,Gonzalez:2017shu,Chabab:2017knz,Rincon:2018sgd,Pani:2009wy,Pani:2012bp,Konoplya:2020bxa,Daghigh:2020mog,Daghigh:2020fmw,Fu:2022cul,Daghigh:2021psm,Fu:2023drp,E:2025kic}. Recent developments have further emphasized the interplay between QNMs, shadows, instability criteria, odd-parity perturbations of hairy geometries, discontinuous effective potentials, and the classification of competing mode families, both in concrete phenomenological models and in more general analyses \cite{Okyay:2021nnh,Pantig:2022gih,Yang:2024cjf,Qian:2024iaq,Li:2025ljb,Xiong:2025wgs}. Recently, greybody factors have attracted renewed attention as observables directly connected with the ringdown signal. In particular, they have been proposed as a way to characterize the frequency-domain amplitude of black-hole ringdown, to provide stable gravitational-wave observables, and to construct new greybody-factor-based tests of strong gravity \cite{Oshita:2023cjz, Rosato:2024arw,Rosato:2026apq}.
This broad literature strongly motivates revisiting the ringdown problem in a hairy black hole generated via gravitational decoupling, where the extra hair arises from an additional anisotropic sector while the spacetime remains asymptotically close to Schwarzschild, making it a natural framework for studying axial spectra, trapping structures, and possible echo-like late-time signatures.

These developments provide strong motivation to investigate black hole solutions that go beyond vacuum general relativity while remaining geometrically and physically well controlled. A particularly useful framework in this direction is gravitational decoupling. In its original formulation, the method was introduced as a systematic way to decouple gravitational sources and generate anisotropic solutions starting from a known seed geometry \cite{Ovalle:2017fgl}. It was later extended to a more general scheme in which both metric functions can be deformed, leading to the extended geometric deformation framework \cite{Ovalle:2018gic}. When this approach is applied to black-hole spacetimes, it gives rise to hairy Schwarzschild-like solutions sustained by an additional anisotropic sector, thereby providing a concrete arena in which black-hole hair can be studied beyond the standard vacuum case \cite{Ovalle:2020kpd}. The logarithmic solution studied here is asymptotically Schwarzschild and is supported by an effective anisotropic source \cite{Avalos:2023ywb}. Its physical interpretation is subject to the energy-condition constraints discussed below.

Gravitational-wave echoes are delayed secondary pulses in the ringdown signal, which can arise from repeated reflections inside an effective cavity produced by near-horizon modifications, exotic compact objects, or double-barrier effective potentials~\cite{Cardoso:2016oxy,Cardoso:2017cqb,Abedi:2016hgu,Mark:2017dnq,Bueno:2017hyj,Konoplya:2018yrp,Li:2019kwa,Yang:2024prm,Tan:2024qij,Duran-Cabaces:2025sly,Zhu:2024gvl,Dong:2020odp,Huang:2021qwe,Liu:2021aqh,Ou:2021efv,Hui:2023ibl,Shen:2024rup,Zhang:2025ygb}.
From the viewpoint of gravitational-wave phenomenology, such hairy backgrounds are interesting not only because they modify the QNM spectrum, but also because they can qualitatively alter the effective potential that governs perturbations. In particular, additional structure in the potential may create trapping regions or cavity-like configurations capable of supporting long-lived excitations and delayed secondary pulses in the time-domain signal \cite{Yang:2024rms,Guo:2022umh,Konoplya:2025uiq}. This raises the intriguing possibility that gravitational-wave echoes may emerge as a dynamical consequence of the background geometry itself, rather than from ad hoc reflective boundary conditions introduced by hand near the horizon. Therefore, understanding whether this mechanism is realized in well-motivated black hole solutions is important both for classical perturbation theory and for the interpretation of possible nonstandard ringdown signatures in present and future LIGO--Virgo observations.

Cavalcanti \textit{et al.} ~\cite{Cavalcanti:2022cga} considered a minimally coupled massless test scalar on an earlier, exponentially deformed gravitational-decoupling hairy black-hole background and determined scalar QNMs and time-domain wave profiles. The present manuscript instead uses the logarithmic geometry and derives the fixed-source odd-parity gravitational master equation. We compute its axial QNMs in the single-barrier sector and identify echo-like signals generated by a double-peak axial potential. Hence, the novelty of the present work lies in the axial gravitational analysis of this logarithmic background and the geometric identification of its axial trapping cavity.

A complementary analysis of a different, exponentially deformed hairy Schwarzschild black hole classified the modes of a double-peak potential into a delocalized photon-sphere family and a localized echo family and studied their time-dependent competition~\cite{Yang:2026qmf}. That work focuses primarily on electromagnetic perturbations and a different metric. It nevertheless provides useful context for interpreting the trapping and echo-like behavior associated with double-peak effective potentials. We do not perform a complete mode-family decomposition or excitation-coefficient analysis; our present claims concern the axial potential morphology and the echo-like response obtained from its time-domain evolution.

This paper is organized as follows. In Sec.~\ref{subsec:GD-short-review}, we review the hairy black hole geometry and discuss the horizon structure. In Sec.~\ref{subsec:axial}, we derive the axial gravitational master equation and the corresponding effective potential. In Sec.~\ref{section:qnm}, we study the QNM spectrum in the single-barrier regime.
In Sec.~\ref{sec:time-domain}, we analyze the time-domain waveforms and the emergence of echo-like signals in the double-peak regime. In Sec.~\ref{light-ring}, we discuss the trapping structure, light rings, and the limitations of the present stability assessment. We conclude in Sec.~\ref{sec:conclusion}.

\section{The hairy black hole via gravitational decoupling}
\label{subsec:GD-short-review}

Gravitational decoupling provides a systematic way to split the total gravitational source into simpler sectors and generate solutions from a known seed geometry \cite{Ovalle:2017fgl}. This construction was subsequently generalized to the extended case, in which both the temporal and radial metric components are deformed \cite{Ovalle:2018gic}. When applied to the Schwarzschild seed, this framework leads to hairy black hole geometries supported by an additional source \cite{Ovalle:2020kpd}. The logarithmic hairy black hole considered in the present work is obtained within this extended gravitational-decoupling framework. For completeness, we briefly summarize the origin of the logarithmic hairy black hole. The extended gravitational-decoupling construction begins with a split of the total source,
\begin{equation}
G_{\mu\nu}=\kappa^2\left(T_{\mu\nu}+\Theta_{\mu\nu}\right),
\label{eq:source-split}
\end{equation}
where $T_{\mu\nu}$ generates a known seed solution, whereas $\Theta_{\mu\nu}$ denotes the additional anisotropic sector. Starting from the Schwarzschild vacuum seed, both metric functions undergo geometric deformations. Requiring the temporal and radial metric functions to share a common horizon, through $e^{\nu}=e^{-\lambda}$, implies $p_r=-\rho$ for the effective source and allows the representation
\begin{equation}
e^{\nu(r)}=\left(1-\frac{2M}{r}\right)H(r).
\end{equation}
The construction of Ref.~\cite{Avalos:2023ywb} closes the auxiliary sector by imposing the weak energy condition. In terms of a positive generating function $\mathcal{G}(r)$, the relevant conditions can be written as
\begin{equation}
1-H-(r-2M)H'=\mathcal{G}(r),
\qquad
2\mathcal{G}(r)-r\mathcal{G}'(r)\geq0.
\end{equation}
Choosing
\begin{equation}
\mathcal{G}(r)=\frac{\alpha M}{r^2}\ln\!\left(\frac{r}{\beta}\right)
\end{equation}
and fixing the integration constant so that $H\to 1$ as $\alpha\to 0$ gives
\begin{equation}
H(r)=1+\frac{\alpha M}{r(r-2M)}
\left[1+\ln\!\left(\frac{r}{\beta}\right)\right].
\end{equation}
Substituting Eq.~(5) into Eq.~(2) yields the static, spherically symmetric hairy black hole metric~\cite{Avalos:2023ywb}
\begin{equation} ds^2=e^{\nu(r)}dt^2-e^{\lambda(r)}dr^2-r^2d\Omega^2 , \label{metric_general} \end{equation} where the metric functions satisfy \begin{equation} e^{\nu(r)}=e^{-\lambda(r)} =1-\frac{2M}{r} +\frac{\alpha M}{r^2} +\frac{\alpha M}{r^2}\ln\left(\frac{r}{\beta}\right). \label{metric_function} \end{equation} Here \(M\) is the black-hole mass, while $\alpha$ and $\beta$ are additional parameters associated with the hairy deformation. For \(\alpha=0\), Eq.~\eqref{metric_function} reduces to the Schwarzschild metric.

Both $\alpha$ and $\beta$ have dimensions of length. In the original construction, $\alpha\geq0$ measures the deformation strength and serves as primary hair not associated with a Gauss-law charge, whereas $\beta$ sets the radial scale entering the logarithm. The effective anisotropic source corresponding to Eq.~\eqref{metric_function} is
\begin{align}
\kappa^2\rho
&=\frac{\alpha M}{r^4}\ln\!\left(\frac{r}{\beta}\right),
&
\kappa^2p_r
&=-\frac{\alpha M}{r^4}\ln\!\left(\frac{r}{\beta}\right),
\\
\kappa^2p_t
&=\frac{\alpha M}{r^4}
\left[\ln\!\left(\frac{r}{\beta}\right)-\frac{1}{2}\right].
\end{align}
Because $p_r=-\rho$, the only nontrivial weak energy condition is $\rho+p_t\geq0$. Since the logarithm increases monotonically outside the event horizon, the weak energy condition holds for all $r\geq r_h$ provided
\begin{equation}
\ln\!\left(\frac{r_h}{\beta}\right)\geq\frac14,
\qquad\text{equivalently}\qquad
\beta\leq r_h e^{-1/4}.
\label{eq:wec-domain}
\end{equation}
The two-parameter scan treats $\alpha$ and $\beta$ as independent geometric parameters and therefore extends beyond the subdomain in which the weak energy condition is satisfied. For the representative parameter choices used in the echo analysis, we find that $\ln(r_h/\beta)<1/4$; thus, these configurations lie outside the weak-energy-condition subfamily of the original matter construction. They should therefore be interpreted as effective geometries for studying the axial trapping and echo mechanism, rather than as configurations that satisfy the weak energy condition throughout the exterior region.

For the metric function above, the horizons are determined by the positive roots of $e^{\nu(r)}=0$, namely
\begin{equation}
r^{2}-2Mr+\alpha M\left(1+\ln\frac{r}{\beta}\right)=0.
\end{equation}
Hence, the number of horizons is entirely determined by the positive-root structure of the transcendental equation above. If there is only one positive root, it corresponds to the unique event horizon $r_h$. If multiple positive roots exist, the largest root defines the outer event horizon, whereas the smaller roots correspond to inner horizons. In particular, in the three-root region, if the positive roots are ordered as
$r_1<r_2<r_3$,
then the event horizon is given by
$r_h=r_3$.

\begin{figure*}[htbp]
\begin{center}
\makebox[\textwidth][c]{
\hspace{-1.0cm}
\includegraphics[width=0.5\textwidth]{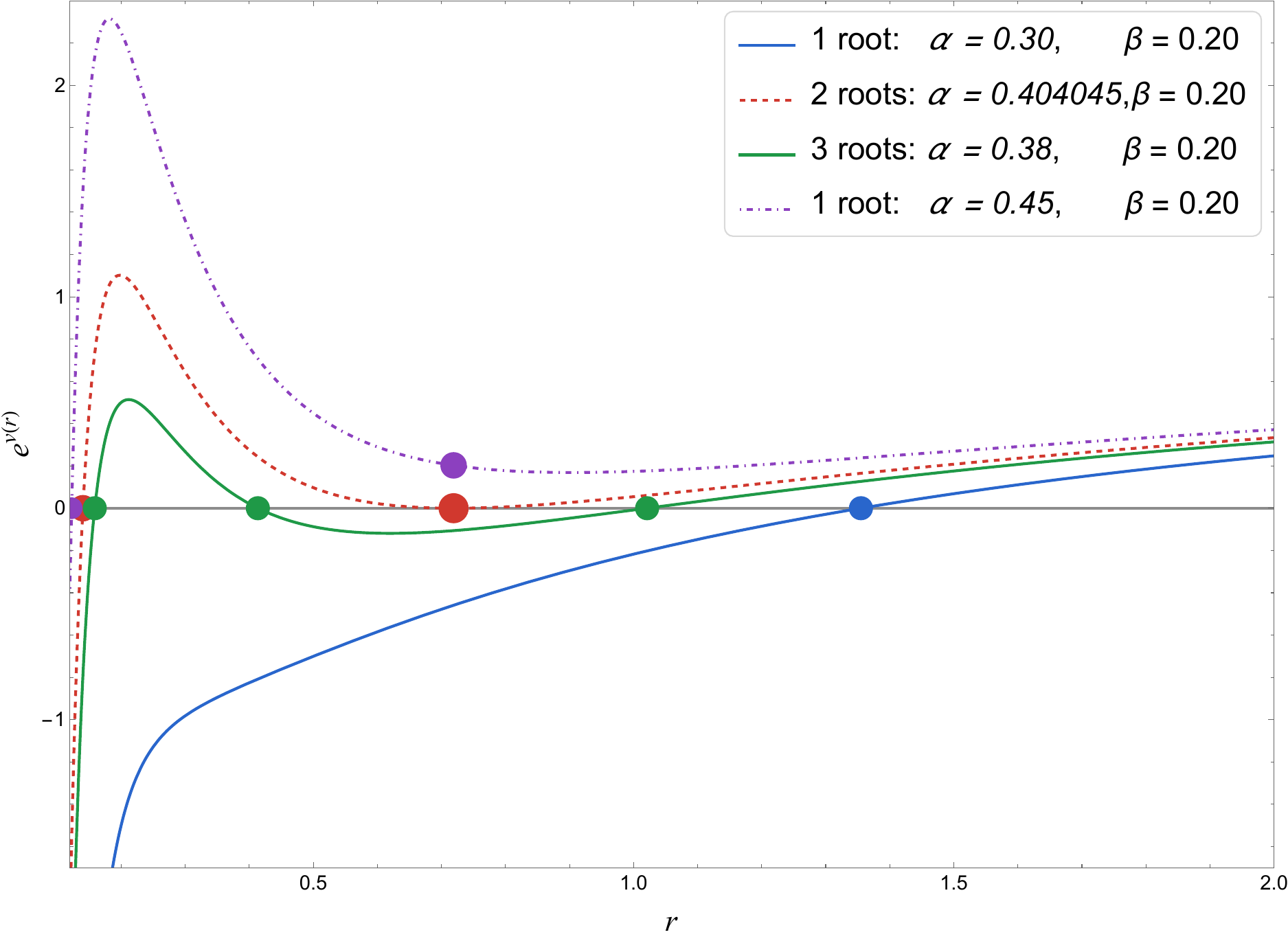}
\hspace{-0.1cm}
\includegraphics[width=0.5\textwidth]{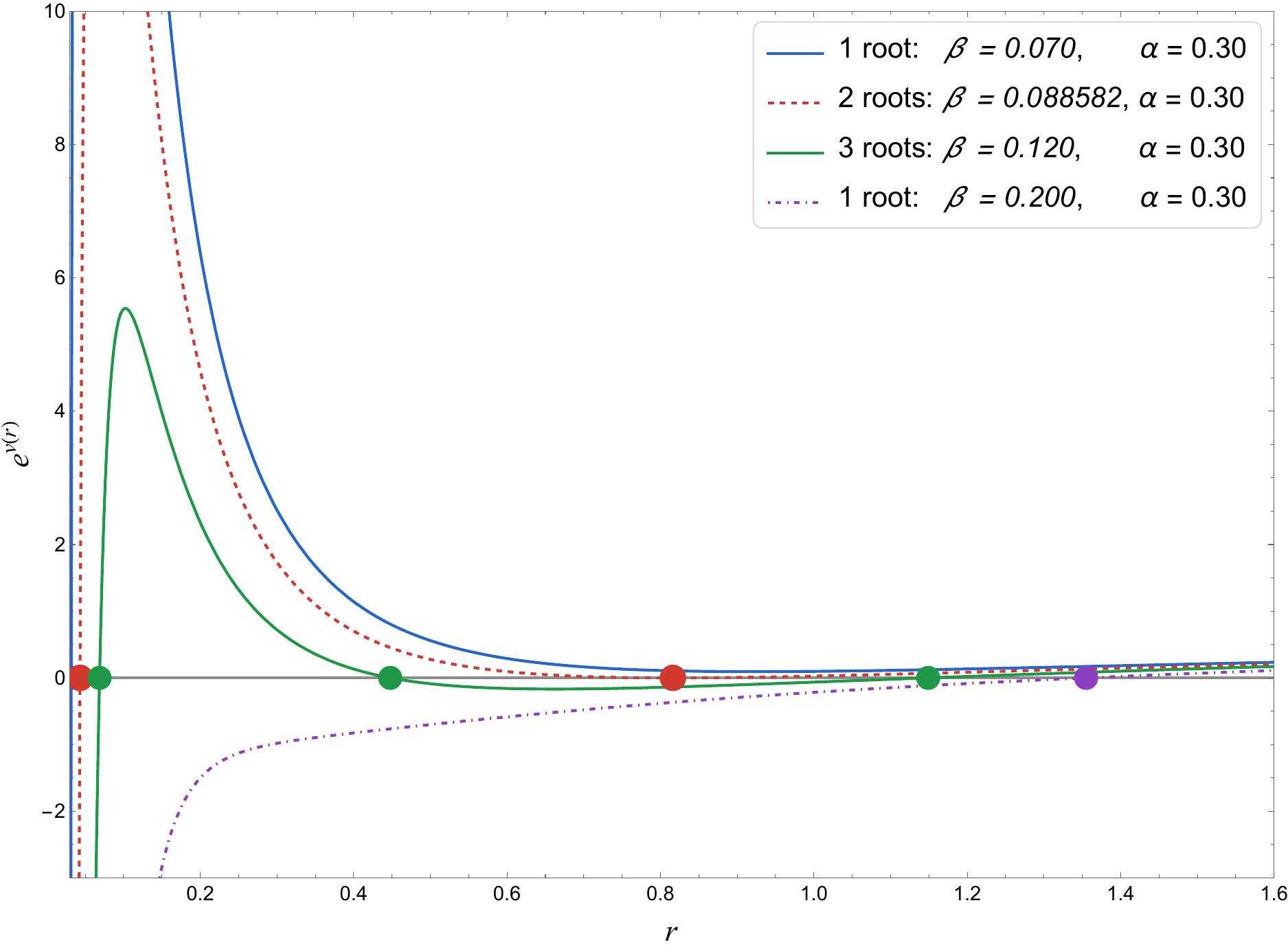}}
\end{center}
\setlength{\abovecaptionskip}{-0.2cm}
\setlength{\belowcaptionskip}{0.5cm}
\caption{The metric function $f(r)$ for several choices of $\alpha$ and $\beta$ with $M=1$. In the left panel, $\beta=0.20$ is fixed and $\alpha$ is varied. In the right panel, $\alpha=0.30$ is fixed and $\beta$ is varied. The blue and purple curves correspond to cases with one positive root. The red curve denotes the extremal case with two positive roots and one degenerate horizon. The green curve shows the case with three distinct positive roots.}
\label{fig:horizon-structure}
\end{figure*}

\begin{figure*}[htbp]
\begin{center}
\makebox[\textwidth][c]{
\hspace{-1.0cm}
\includegraphics[width=0.495\textwidth]{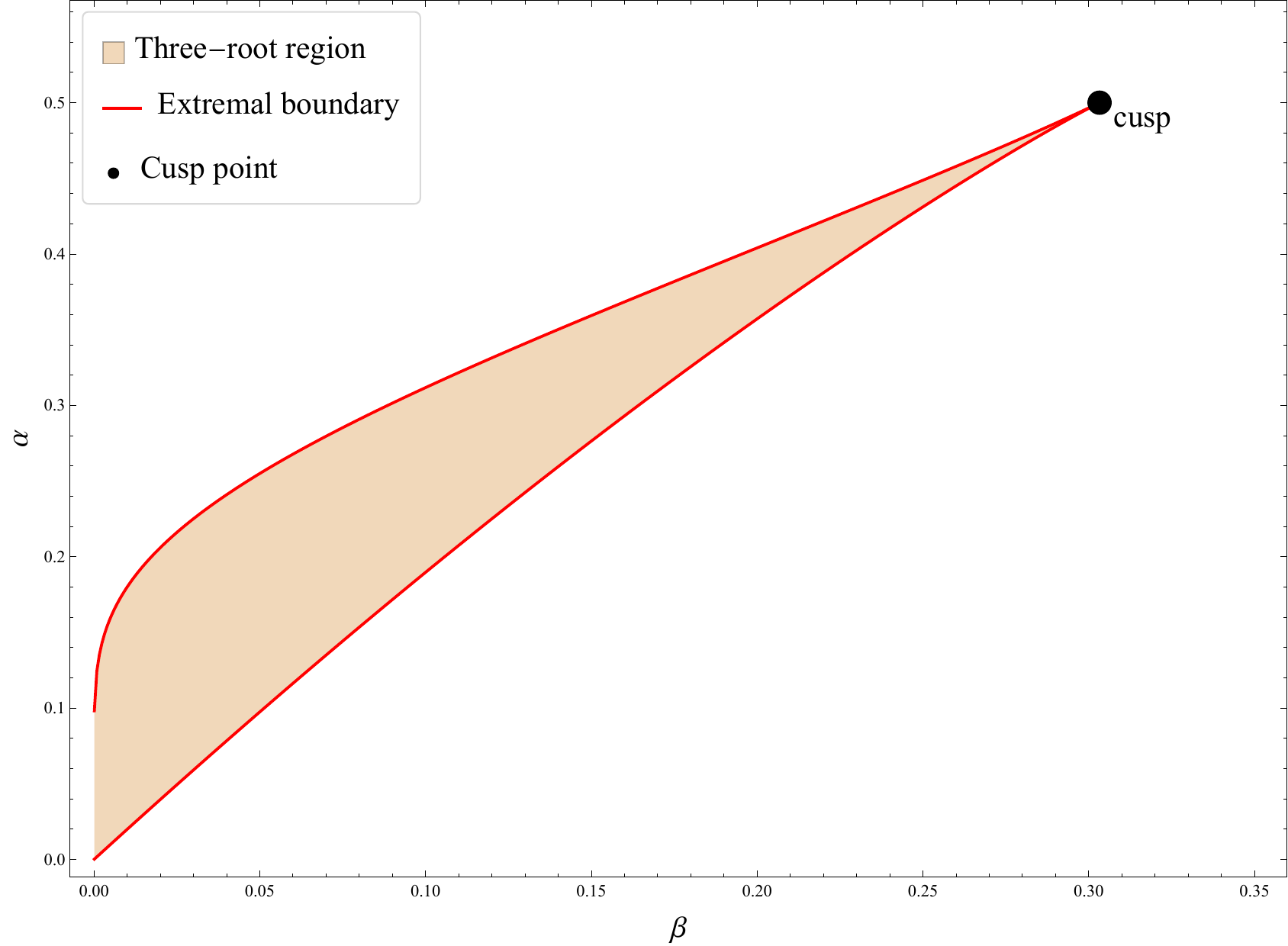}}
\end{center}
\setlength{\abovecaptionskip}{-0.2cm}
\setlength{\belowcaptionskip}{0.8cm}
\caption{
Parameter-space structure in the $(\beta,\alpha)$ plane for the hairy black hole with $M=1$. The shaded region indicates the parameter range with three distinct positive roots. The red curves represent the extremal boundaries, where two positive roots merge into a degenerate horizon. Outside the shaded region, the metric function has a single positive root.}
\label{alpha_beta}
\end{figure*}

Fig.~\ref{fig:horizon-structure} shows the different root structures of the metric function \(e^{\nu(r)}\) as the parameters vary. In the left panel, \(\beta=0.20\) is fixed and \(\alpha\) varies, whereas in the right panel, \(\alpha=0.30\) is fixed and \(\beta\) varies. The blue and purple curves each have one positive root, the red curve corresponds to the extremal case with a degenerate horizon, and the green curve has three distinct positive roots. This behavior can also be understood from the parameter-space structure shown in Fig.~\ref{alpha_beta}. The beige shaded region in the \((\beta,\alpha)\) plane corresponds to the parameter range in which the metric function admits three distinct positive roots, whereas outside this region it has only one positive root. The two red boundary curves represent extremal configurations, for which two roots merge into a degenerate horizon \(r_e\). To determine these extremal boundaries, one imposes the double-root conditions. The extremal boundary is therefore determined by
\begin{equation}
e^{\nu(r)}=0,~ \left(e^{\nu(r)}\right)'=0 .
\end{equation}
For the present metric, one has
\begin{equation}
f'(r)=\frac{M}{r^{3}}\left[2r-\alpha\left(1+2\ln\frac{r}{\beta}\right)\right],
\end{equation}
so that the extremal conditions become
\begin{equation}
r_e^{2}-2Mr_e+\alpha M\left(1+\ln\frac{r_e}{\beta}\right)=0,
\end{equation}
and
\begin{equation}
2r_e-\alpha\left(1+2\ln\frac{r_e}{\beta}\right)=0.
\end{equation}

After introducing the dimensionless quantities,
\begin{equation}
\bar r_e=\frac{r_e}{M},\qquad
\bar\alpha=\frac{\alpha}{M},\qquad
\bar\beta=\frac{\beta}{M},
\end{equation}
the extremal boundary can be parameterized as
\begin{equation}
\bar\alpha=2\bar r_e(1-\bar r_e),\qquad
\bar\beta=\bar r_e\exp\!\left[-\frac{\bar r_e}{2(1-\bar r_e)}\right].
\end{equation}
The two branches meet at the cusp point
\begin{equation}
(\bar\beta,\bar\alpha)=\left(\frac{1}{2\sqrt e},\frac12\right).
\end{equation}

\section{Axial perturbations of the gravitational field for the hairy black hole}
\label{subsec:axial}

We now consider odd-parity gravitational perturbations of the hairy black-hole background. The effective matter sector supporting this background may be represented as an anisotropic fluid, 
\begin{equation} 
T^\mu_{\ \nu} = {\rm diag}\left(\rho,-p_r,-p_t,-p_t\right). 
\end{equation} 
In a complete perturbative treatment of an anisotropic-fluid background, the
odd-parity sector may, in general, contain axial matter perturbations, such as
axial velocity perturbations or odd-parity anisotropic-stress perturbations.
For instance, the odd-parity perturbation sector of electrically charged Kalb--Ramond black holes is characterized by a coupled gravitational--electromagnetic system \cite{Gu2026}.
For the hairy geometry considered here, however, the additional source is an
effective anisotropic sector introduced through gravitational decoupling. No
independent microscopic action or equation of state is specified for this
source. Therefore, a full coupled treatment of metric and matter perturbations
would require additional assumptions regarding the dynamics of the effective
matter sector.

In the present work, we adopt a fixed-source axial perturbation scheme. In this
scheme, the odd-parity response of the effective matter sector is neglected,
and we set
\begin{equation}
\delta T_{\mu\nu}^{\rm (ax)}=0 .
\end{equation}
The resulting perturbation equation should therefore be understood as the axial equation for the fixed hairy background, rather than as the most general axial perturbation equation of the full metric--matter system.
This fixed-source assumption is central to the scope of the present results; it does not demonstrate that the effective matter sector is dynamically inert. Our calculation tests the reduced odd-parity metric operator on the prescribed background. It does not establish the stability of a microscopic completion of $\Theta_{\mu\nu}$, nor does it cover polar perturbations, for which density, pressure, and anisotropic-stress perturbations may enter explicitly.

Having specified the scope of the fixed-source approximation, we now turn to the axial perturbation equations.
The linearized field equations are \begin{equation} 
\delta G_{\mu\nu}=8\pi \delta T_{\mu\nu}^{\rm (ax)} . 
\end{equation} 
Under the fixed-source assumption above, they reduce to 
\begin{equation} 
\delta G_{\mu\nu}^{\rm (ax)}=0 . 
\end{equation} 
In the Regge--Wheeler gauge, the odd-parity metric perturbation is written as
\begin{equation}
h_{\mu\nu}^{\rm odd}
=
\left(
\begin{array}{cccc}
0 & 0 & 0 & h_0(t,r) \\
0 & 0 & 0 & h_1(t,r) \\
0 & 0 & 0 & 0 \\
h_0(t,r) & h_1(t,r) & 0 & 0
\end{array}
\right)
\sin\theta\,\partial_\theta P_\ell(\cos\theta),
\label{axial_metric_perturbation}
\end{equation}
where \(P_\ell(\cos\theta)\) is the Legendre polynomial. The functions
\(h_0(t,r)\) and \(h_1(t,r)\) describe the odd-parity perturbation. We assume the
harmonic time dependence
\begin{equation}
h_0(t,r)=h_0(r)e^{-i\omega t},
\qquad
h_1(t,r)=h_1(r)e^{-i\omega t}.
\end{equation}

Substituting Eq.~\eqref{axial_metric_perturbation} into the linearized field equations, the nontrivial axial equations reduce to two coupled radial equations,
\begin{equation}
h_1(r)
\left[
r^2\omega^2-(\ell-1)(\ell+2)e^{\nu(r)}
\right]
-ir^2\omega h_0'(r)
+2ir\omega h_0(r)
=0,
\label{axial_eq_1}
\end{equation}
and
\begin{equation}
e^{\nu(r)}
\left[
\frac{h_1(r)\left(e^{\nu(r)}\right)'}{e^{\nu(r)}}
+2h_1'(r)
\right]
+
\frac{2i\omega h_0(r)}{e^{\nu(r)}}
=0.
\label{axial_eq_2}
\end{equation}
Here, a prime denotes differentiation with respect to \(r\).

To derive the master equation, we introduce the tortoise coordinate
\begin{equation}
\frac{dr_*}{dr}=e^{-\nu(r)},
\label{tortoise_coordinate}
\end{equation}
and define the master variable by
\begin{equation}
h_1(r)=r e^{-\nu(r)}\Psi(r).
\label{master_variable}
\end{equation}
We then combine Eqs.~\eqref{axial_eq_1} and \eqref{axial_eq_2} to eliminate \(h_0(r)\) and \(h_1(r)\). As a result, the axial perturbations are described by a single master variable \(\Psi\). Therefore, the axial perturbation equation can be written as
\begin{equation}
\frac{\partial^2\Psi(t,r)}{\partial r_*^2}
-
\frac{\partial^2\Psi(t,r)}{\partial t^2}
-
V_{\rm ax}(r)\Psi(t,r)
=0,
\label{axial_master_equation_time}
\end{equation}
where the axial effective potential is defined as
\begin{equation}
V_{\rm ax}(r)
=
e^{\nu(r)}
\left[
\frac{\ell(\ell+1)}{r^2}
-\frac{6M}{r^3}
+\frac{\alpha M}{r^4}
\left(
3+4\ln\left(\frac{r}{\beta}\right)
\right)
\right].
\label{axial_potential}
\end{equation}
When \(\alpha=0\), the above potential reduces to the standard Regge--Wheeler potential of the Schwarzschild black hole.
\begin{equation}
V_{\rm Sch}(r)
=
\left(1-\frac{2M}{r}\right)
\left[
\frac{\ell(\ell+1)}{r^2}
-\frac{6M}{r^3}
\right].
\end{equation}

With the axial master equation and the corresponding effective potential established, one can proceed to study the QNM spectrum and the time-domain evolution of the perturbation field.
These results are presented in the following sections.

\section{QNMs of the hairy black hole} \label{section:qnm}
In this section, we study the QNM spectrum of the hairy black hole spacetime. 
We compute the QNM frequencies and analyze the corresponding quasinormal ringdown signals. 
The results are obtained using the pseudospectral method, the WKB method, and Prony extraction from the time-domain evolution. 
We then compare these results and discuss their dependence on the parameter $\alpha$. 
This analysis helps reveal how changes in the spacetime geometry affect the oscillation and damping properties of the axial gravitational perturbations.

The pseudospectral method \cite{Konoplya:2024lch,Jaramillo:2020tuu,Jansen:2017oag} is a powerful tool for solving the perturbation equation and extracting quasinormal frequencies. 
The main idea is to discretize the master equation on a set of Chebyshev collocation points and transform the differential equation into a matrix eigenvalue problem. 
The QNM frequencies are then obtained from the resulting eigenvalues.

For the present hairy black hole, we implement the pseudospectral approach using a compactified hyperboloidal Chebyshev collocation scheme. With the time dependence
$\Psi(t,r)=e^{-i\omega t}\psi(r)$, the standard QNM boundary conditions
are
\begin{equation}
\psi(r)\sim e^{-i\omega r_*},
\qquad r\rightarrow r_h,
\qquad
\psi(r)\sim e^{+i\omega r_*},
\qquad r\rightarrow\infty,
\label{eq:qnm_bc}
\end{equation}
corresponding to a purely ingoing wave at the event horizon and a purely outgoing wave at infinity, respectively. In the present hyperboloidal formulation, the ingoing and outgoing asymptotic behaviors are incorporated into the transformed radial operator; therefore, no separate endpoint boundary conditions are imposed.

The exterior region $r\in[r_h,\infty)$ is compactified by introducing the dimensionless radial coordinate
\begin{equation}
\sigma=\frac{r_h}{r},
\qquad
\sigma\in[0,1],
\label{eq:compactification}
\end{equation}
where $\sigma=1$ corresponds to the event horizon and $\sigma=0$
denotes the asymptotic boundary $r\rightarrow\infty$. Here, $r_h$
is the largest positive root of $f(r_h)=0$. For numerical convenience, we introduce
$q_{\rm ax}(r)=\ell(\ell+1)+2[f(r)-1]-rf'(r)$,
which is related to the axial effective potential.
The auxiliary functions used in the numerical code are
\begin{align}
p(\sigma)
&=
\sigma^2 f\left(\frac{r_h}{\sigma}\right),
\\
\gamma(\sigma)
&=
-1+
\frac{2(r_h+2M\sigma)}{r_h}
f\left(\frac{r_h}{\sigma}\right),
\\
w(\sigma)
&=
\frac{1-\gamma(\sigma)^2}{p(\sigma)}.
\end{align}
In particular, $\gamma\rightarrow-1$ at the event horizon and $\gamma\rightarrow+1$ at infinity, thereby encoding the appropriate ingoing and outgoing characteristic behavior. The resulting equation for the regularized radial function $\phi(\sigma)$ can be written as
\begin{equation}
C\phi''
+
E\phi'
+
\mathcal{W}\phi
-i\widehat{\omega}
\left(A\phi'+B\phi\right)
+
\widehat{\omega}^{\,2}\phi
=0,
\label{eq:hyperboloidal_equation}
\end{equation}
where $\widehat{\omega}=r_h\omega$ and
\begin{equation}
C=\frac{p}{w},
\qquad
E=\frac{p'}{w},
\qquad
A=\frac{2\gamma}{w},
\qquad
B=\frac{\gamma'}{w},
\qquad
\mathcal{W}=-\frac{q_{\rm ax}}{w}.
\end{equation}

The compactified equation is discretized at the interior Chebyshev--Gauss collocation points.
\begin{equation}
\sigma_j=
\frac{1}{2}
\left[
1-\cos\left(\frac{(2j-1)\pi}{2N}\right)
\right],
\qquad
j=1,2,\ldots,N.
\label{eq:chebyshev_points}
\end{equation}
These points lie strictly within the interval $(0,1)$, thereby avoiding direct numerical evaluation at the horizon and at infinity. The derivatives are represented by spectral differentiation matrices,
\begin{equation}
\phi'\rightarrow D^{(1)}\boldsymbol{\phi},
\qquad
\phi''\rightarrow D^{(2)}\boldsymbol{\phi},
\qquad
D^{(2)}=D^{(1)}D^{(1)}.
\end{equation}
The radial equation is then reduced to a quadratic matrix eigenvalue problem.
\begin{equation}
\left[
L_1-i\widehat{\omega}L_2
+\widehat{\omega}^{\,2}I
\right]\boldsymbol{\phi}=0,
\label{eq:quadratic_eigenvalue}
\end{equation}
which is linearized and solved numerically for
$\widehat{\omega}$. The physical frequency is then recovered from
$\omega=\widehat{\omega}/r_h$.

To remove resolution-dependent numerical modes, the spectra obtained
at two nearby resolutions are compared. An eigenvalue
$\hat{\omega}\in\hat{\Lambda}_{N_2}$ is retained as a stable candidate
when
\begin{equation}
\min_{\hat{\mu}\in\hat{\Lambda}_{N_1}}
|\hat{\omega}-\hat{\mu}|
\leq
\epsilon(1+|\hat{\omega}|),
\end{equation}
where $\epsilon$ denotes a prescribed relative tolerance. In practice, this convergence test is performed using nearby spectral resolutions, typically in the range $N\simeq60$--$70$.

Moreover, the QNM frequencies are also computed using the higher-order WKB method.
This method is applicable when the effective potential forms a single barrier with two turning points.
Its basic idea is to match the asymptotic WKB solutions to the local Taylor expansion of the wave function near the maximum of the effective potential.
In this way, the quasinormal frequency can be expressed in terms of the value of the potential and its derivatives at the peak.
Following the standard higher-order WKB approach, the frequency is determined from \cite{Konoplya:2019hlu,Iyer:1986np,Matyjasek:2017psv}
\begin{equation}
\begin{aligned}
\omega^2 ={}& V_0 + A_2(K^2)+A_4(K^2)+A_6(K^2)+\cdots \\
& -iK\sqrt{-2V_2}\left[1+A_3(K^2)+A_5(K^2)+A_7(K^2)+\cdots\right],
\end{aligned}
\end{equation}
where
\begin{equation}
K=n+\frac{1}{2}, \qquad n=0,1,2,\ldots,
\end{equation}
$V_0$ is the value of the effective potential at its maximum, and $V_2$ denotes the second derivative of the potential with respect to the tortoise coordinate at the same point.
The correction terms $A_k$ depend on higher derivatives of the potential at the peak.
To improve the accuracy of the calculation, Pad\'e approximants are further employed in the WKB expansion.

To obtain the time-domain profiles, we use double-null coordinates
\begin{equation}
u=t-r_*, \qquad v=t+r_*.
\end{equation}
Under these conditions, the perturbation equation takes the form
\begin{equation}
4\frac{\partial^2\Psi}{\partial u\,\partial v}
+
V_{\rm ax}(r)\Psi
=
0.
\label{eq:uv-eq}
\end{equation}
A standard discretization scheme on the null grid is given by\cite{Gundlach:1993tp}
\begin{equation}
    \Psi_N = \Psi_W + \Psi_E - \Psi_S
    -\frac{\Delta^2}{8}V_S\left(\Psi_W+\Psi_E\right)
    +\mathcal{O}(\Delta^4),
    \label{eq:fd-scheme}
\end{equation}
where $N$, $W$, $E$, and $S$ denote the usual north, west, east, and south points of the null grid.

The QNM frequencies can also be extracted from the time-domain waveform using the Prony method.
In this method, the waveform is fitted by a finite sum of damped exponentials \cite{Konoplya:2011qq,Nollert:1999ji}.
\begin{equation}
\Psi(t) \simeq \sum_{k=1}^{p} C_k e^{-i\omega_k t},
\end{equation}
where $C_k$ are fitting coefficients and $\omega_k$ are the complex frequencies to be extracted. This method is useful for extracting the quasinormal modes directly from the numerical waveform and for checking the consistency of the results obtained from different approaches.

Table~\ref{qnf-alpha} lists the fundamental QNM frequencies obtained from the pseudospectral method, the WKB approximation, and the Prony method for different values of $\alpha$, with $M=1$, $\ell=2$, and $\beta=0.2$. As a benchmark, we also include the Schwarzschild limit, $\alpha=0$, in Table~\ref{qnf-alpha}. In this limit, the parameter $\beta$ drops out, and the axial effective potential reduces to the standard Regge--Wheeler potential. For $M=1$, $\ell=2$, and $n=0$, the QNM frequency of the Schwarzschild black hole is $\omega=0.3736716844-0.0889623157i$~\cite{Berti:2009kk}. Our results show good agreement with this reference value and thus provide a useful consistency check for the three numerical implementations. As $\alpha$ increases, the real part of the frequency increases monotonically, indicating a higher oscillation frequency. By contrast, the magnitude of the imaginary part exhibits non-monotonic behavior as $\alpha$ increases. It first grows slightly at small values of $\alpha$ and then decreases in the large-$\alpha$ region. This indicates that the damping rate is initially enhanced, but becomes weaker when $\alpha$ is sufficiently large. It is worth noting that the standard WKB approximation is constructed from a local expansion around a single maximum of the effective potential. Therefore, we apply the WKB method only to the single-barrier configurations listed in Table~\ref{qnf-alpha} and do not use it in the double-peak regime.

\begin{figure*}[t]
\begin{center}
\vspace{1cm}
\makebox[\textwidth][c]{
\hspace{-1.0cm}
\includegraphics[width=0.55\textwidth]{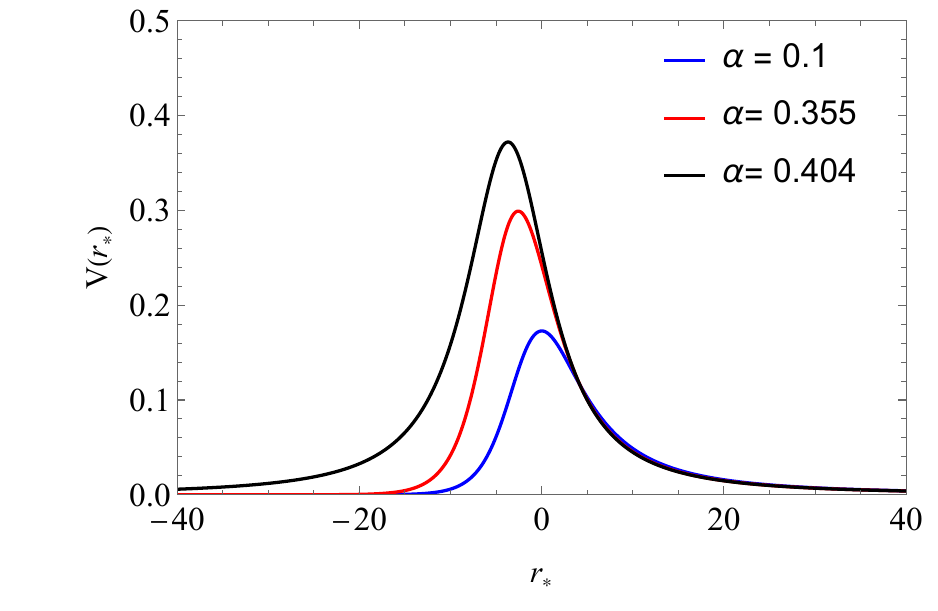}
\hspace{-1.1cm}
\includegraphics[width=0.55\textwidth]{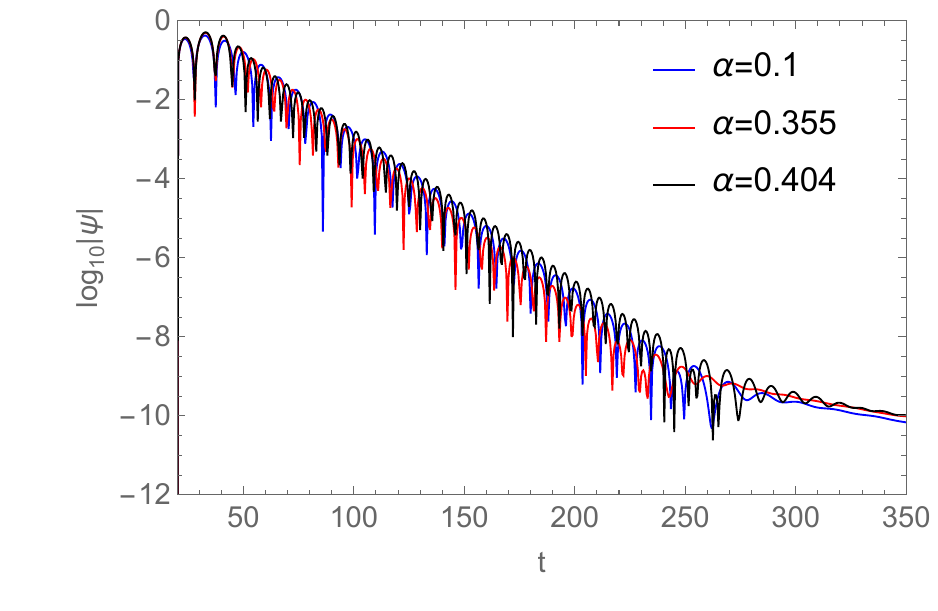}}
\end{center}
\setlength{\abovecaptionskip}{-0.2cm}
\setlength{\belowcaptionskip}{0.5cm}
\caption{Effective potential $V(r_*)$ (left panel) and semilogarithmic time-domain profiles (right panel) for axial gravitational perturbations with $M=1$, $\ell=2$, and $\beta=0.2$. The three curves correspond to $\alpha = 0.1$, $0.355$, and $0.404$.
}
\label{V-alpha}
\end{figure*}

Fig.~\ref{V-alpha} shows the effective potential and the time-domain evolution for three values of $\alpha$. In the left panel, the peak of the effective potential becomes higher as $\alpha$ increases. At the same time, the peak moves to smaller $r_*$. The potential barrier is therefore stronger for larger $\alpha$, which means that the wave experiences stronger trapping near the peak region. The right panel shows the semilogarithmic time-domain profiles. The three curves exhibit similar overall behavior. They all show a clear damped-oscillation stage. The oscillation frequency increases with $\alpha$, which agrees with the larger real part of the quasinormal frequency. The decay rate also changes with $\alpha$, but the trend is not monotonic over the entire parameter range. In particular, the signal for $\alpha=0.404$ decays more slowly than that for $\alpha=0.1$, while the $\alpha=0.355$ case lies in between. This is consistent with the quasinormal frequencies, for which the magnitude of the imaginary part first increases slightly and then decreases in the large-$\alpha$ region. The non-monotonic behavior of the damping rate may be qualitatively related to changes in the horizon structure. For $\beta=0.2$, the three-root region is bounded by $\alpha \approx 0.357164$ and $\alpha \approx 0.404045$. However, the turning point of $|\mathrm{Im}\,\omega|$ appears before the lower boundary of this region. The damping rate starts to decrease slightly before the lower boundary of the three-root region, and this decrease becomes more pronounced as $\alpha$ approaches and enters that region. This suggests that the change in the horizon structure in the large-$\alpha$ regime may be associated with the weakening of the damping, although it is not the sole origin of the effect.

\begin{table}[t]
\centering
\caption{Fundamental QNM frequencies for different values of $\alpha$ with $M=1$, $\ell=2$, and $\beta=0.2$.}
\renewcommand{\arraystretch}{1.6}
\setlength{\tabcolsep}{18pt}
\begin{tabular}{cccc}
\hline\hline
$\alpha$ & $\omega$ (Pseudospectral) & $\omega$ (WKB) & $\omega$ (Prony) \\
\hline\hline
0.0      & $0.373672 - 0.0889623\,i$ & $0.373619 - 0.0888910\,i$ & $0.373822 - 0.0887893\,i$ \\
0.1      & $0.400464 - 0.0922784\,i$ & $0.400516 - 0.0922427\,i$ & $0.400684 - 0.0921576\,i$ \\
0.2      & $0.436263 - 0.0957293\,i$ & $0.436360 - 0.0957928\,i$ & $0.436604 - 0.0956848\,i$ \\
0.3      & $0.489404 - 0.0984235\,i$ & $0.489442 - 0.0986107\,i$ & $0.489779 - 0.0983139\,i$ \\
0.34     & $0.520374 - 0.0983334\,i$ & $0.520299 - 0.0985215\,i$ & $0.520596 - 0.0981727\,i$ \\
0.355    & $0.534554 - 0.0977710\,i$ & $0.534425 - 0.0979225\,i$ & $0.534711 - 0.0976157\,i$ \\
0.357164 & $0.536753 - 0.0976494\,i$ & $0.536617 - 0.0977923\,i$ & $0.536898 - 0.0974844\,i$ \\
0.365    & $0.545084 - 0.0970954\,i$ & $0.544925 - 0.0971991\,i$ & $0.545425 - 0.0970263\,i$ \\
0.38     & $0.562929 - 0.0953301\,i$ & $0.562759 - 0.0953165\,i$ & $0.563015 - 0.0950163\,i$ \\
0.395    & $0.583944 - 0.091793\,i$ & $0.583915 - 0.0916647\,i$ & $0.584504 - 0.0913758\,i$ \\
0.404    & $0.598674 - 0.0876769\,i$ & $0.598652 - 0.0877838\,i$ & $0.599320 - 0.0875141\,i$ \\
\hline\hline
\end{tabular}
\vspace{1.5cm}
\label{qnf-alpha}
\end{table}

Notably, for $\beta=0.2$, when $\alpha$ exceeds the upper boundary value $\alpha \approx 0.404045$, the effective potential develops a double-peak structure. This structure is a necessary condition for the appearance of echo-like signals, as it creates a cavity between the two barriers and permits repeated partial reflections of the wave. Therefore, the region above this boundary provides a natural parameter window in which one may search for gravitational-wave echoes.

\section{Echoes of the hairy black hole spacetime}

Motivated by the discussion above, we now turn to the echo properties of the hairy black hole spacetime. Our main focus is the parameter region in which the axial effective potential develops a double-peak structure. After a systematic numerical scan, we find that such a structure indeed appears in part of the upper one-root branch above the three-root region. For each pair $(\alpha,\beta)$, we first determine the event horizon $r_h$ and then examine the axial effective potential $V_{\rm ax}(r_*)$ in the exterior region. A configuration is classified as belonging to the double-peak region if three distinct extrema $r_{*L}<r_{*m}<r_{*R}$ exist and satisfy
\begin{equation}
\left.
\frac{dV_{\rm ax}}{dr_*}
\right|_{r_{*L}}
=
\left.
\frac{dV_{\rm ax}}{dr_*}
\right|_{r_{*m}}
=
\left.
\frac{dV_{\rm ax}}{dr_*}
\right|_{r_{*R}}
=0,
\end{equation}
with
\begin{equation}
\left.
\frac{d^2V_{\rm ax}}{dr_*^2}
\right|_{r_{*L}}<0,
\qquad
\left.
\frac{d^2V_{\rm ax}}{dr_*^2}
\right|_{r_{*m}}>0,
\qquad
\left.
\frac{d^2V_{\rm ax}}{dr_*^2}
\right|_{r_{*R}}<0.
\end{equation}
Thus, $r_{*L}$ and $r_{*R}$ correspond to the two potential barriers, while $r_{*m}$ denotes the local minimum between them, thereby forming an effective trapping cavity. We therefore analyze the time-domain signal in this sector and examine how the delay, amplitude, and repetition pattern of the late-time pulses depend on the shape of this effective cavity. In the discussion below, we refer to this domain as the echo-supporting branch of the solution. 
\label{sec:time-domain}
\begin{figure*}[b]
\begin{center}
\makebox[\textwidth][c]{
\hspace{-1.0cm}
\includegraphics[width=0.55\textwidth]{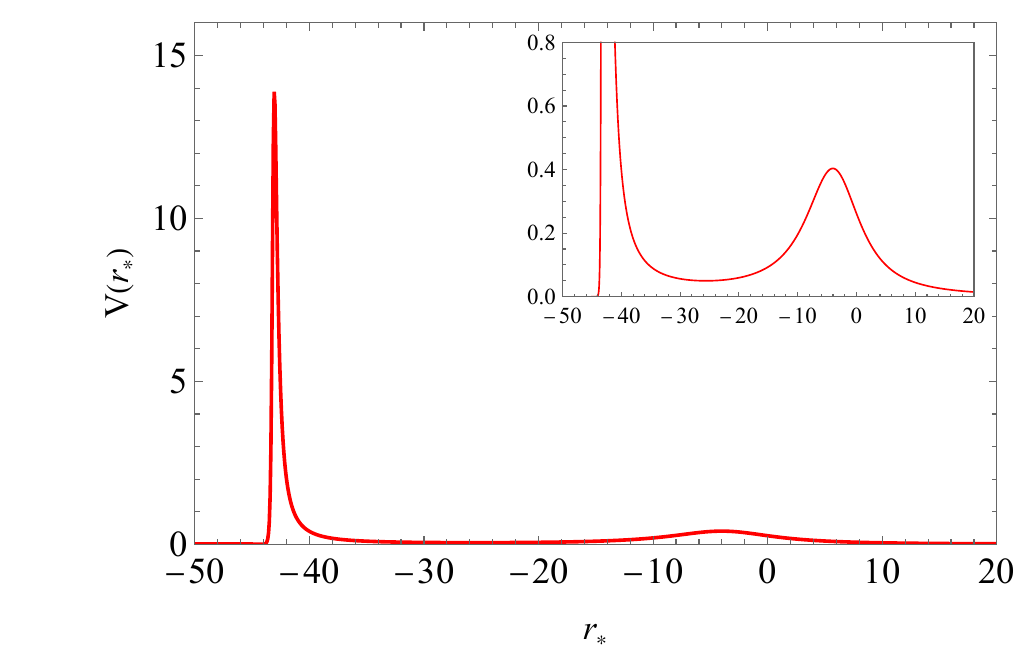}
\hspace{-1.1cm}
\includegraphics[width=0.55\textwidth]{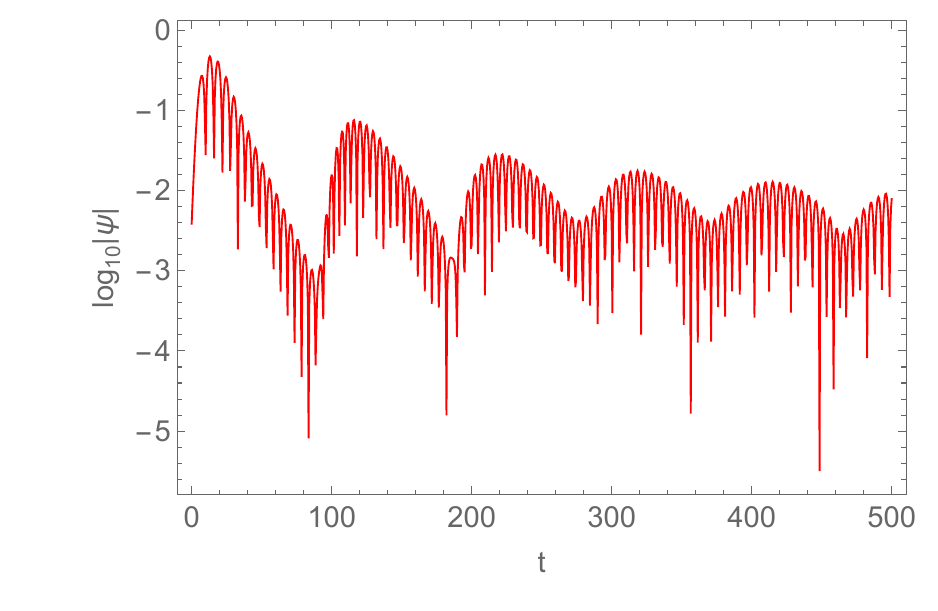}}
\makebox[\textwidth][c]{
\hspace{-1.0cm}
\includegraphics[width=0.55\textwidth]{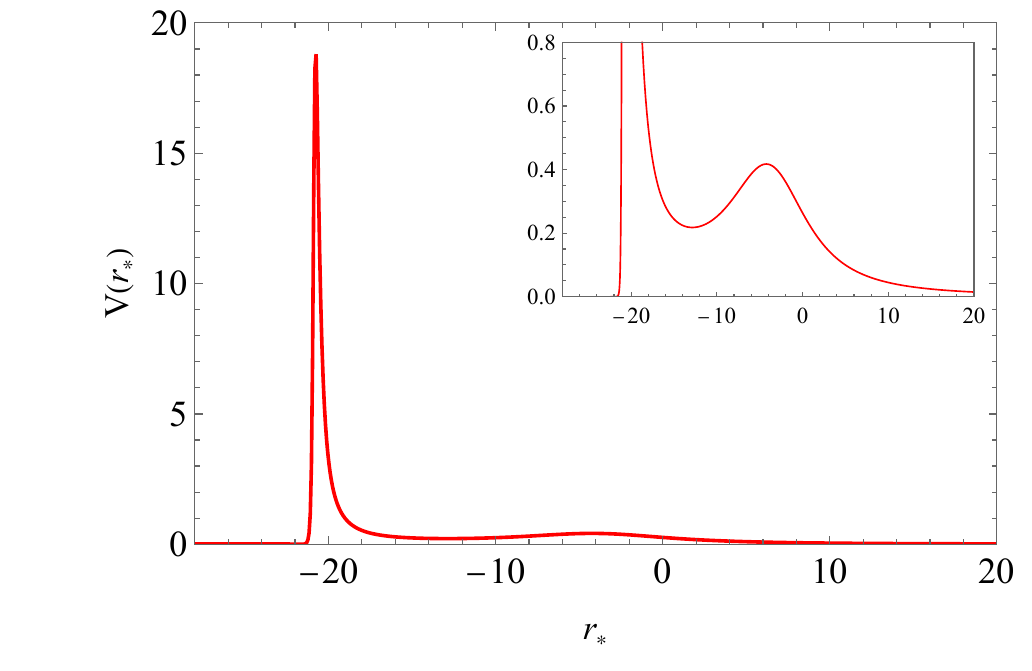}
\hspace{-1.1cm}
\includegraphics[width=0.55\textwidth]{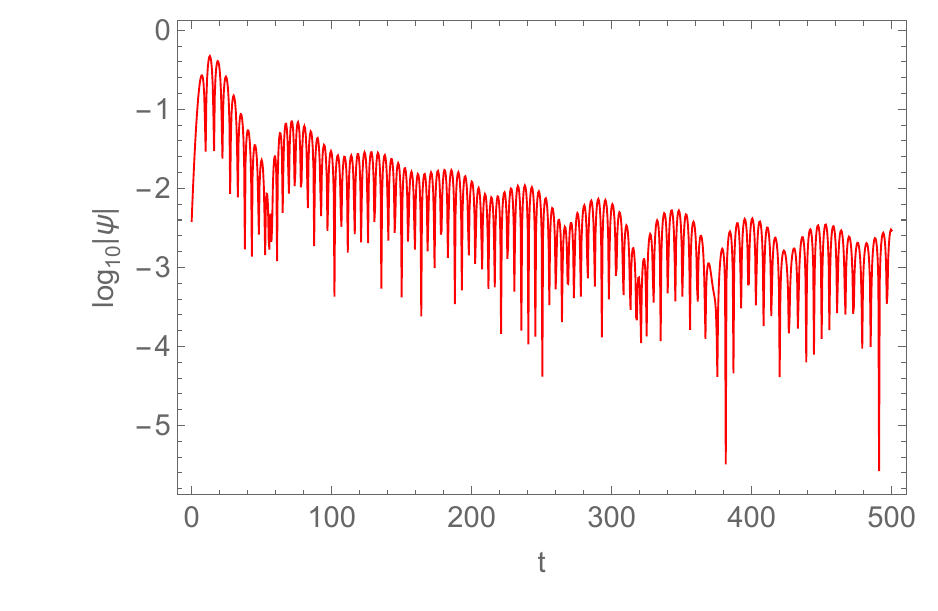}}
\makebox[\textwidth][c]{
\hspace{-1.0cm}
\includegraphics[width=0.55\textwidth]{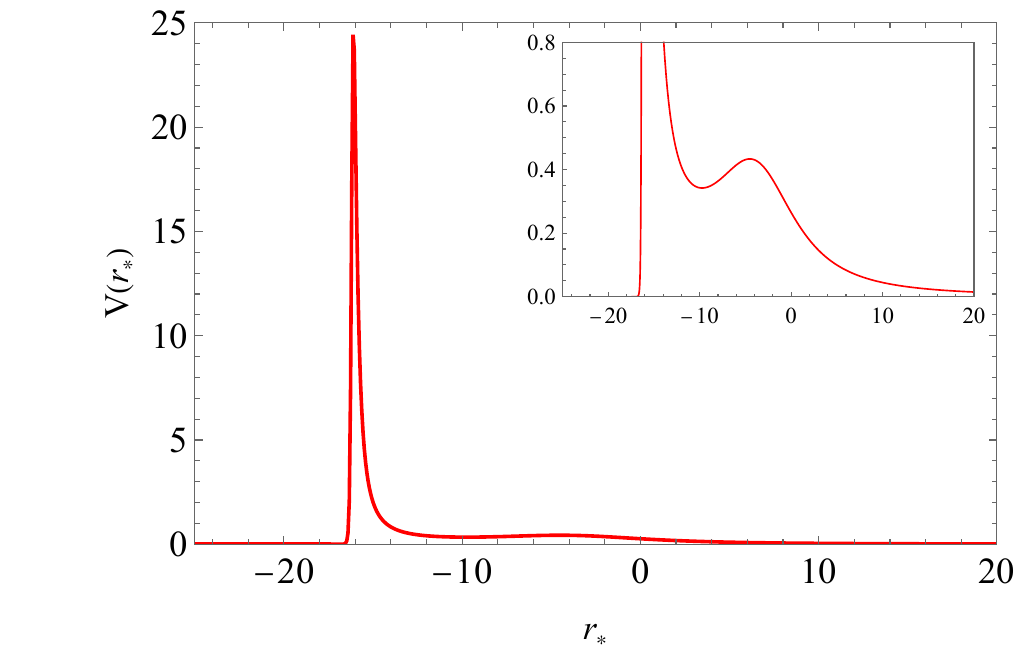}
\hspace{-1.1cm}
\includegraphics[width=0.55\textwidth]{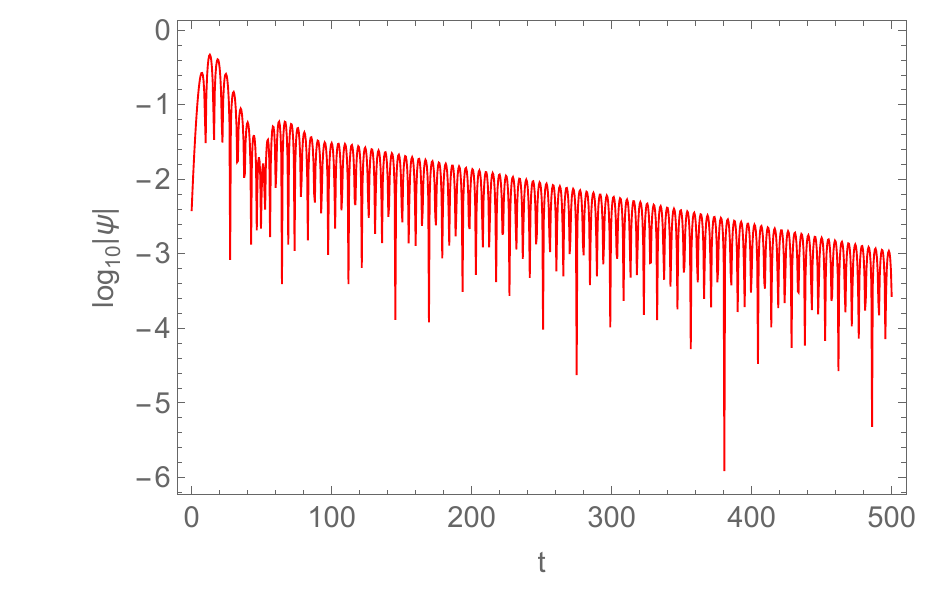}}
\end{center}
\setlength{\abovecaptionskip}{-0.2cm}
\setlength{\belowcaptionskip}{0.5cm}
\caption{Effective potential as a function of the tortoise coordinate $r_*$ (left panel) and the corresponding time-domain waveform as a function of time $t$ (right panel). The parameters are fixed at $M=1$, $\ell=2$, and $\beta=0.25$. From top to bottom, the three cases correspond to $\alpha=0.45$, $0.455$, and $0.46$, respectively.
}
\label{echo-beta025}
\end{figure*}

\begin{figure*}[htbp]
\begin{center}
\makebox[\textwidth][c]{
\hspace{-1.0cm}
\includegraphics[width=0.55\textwidth]{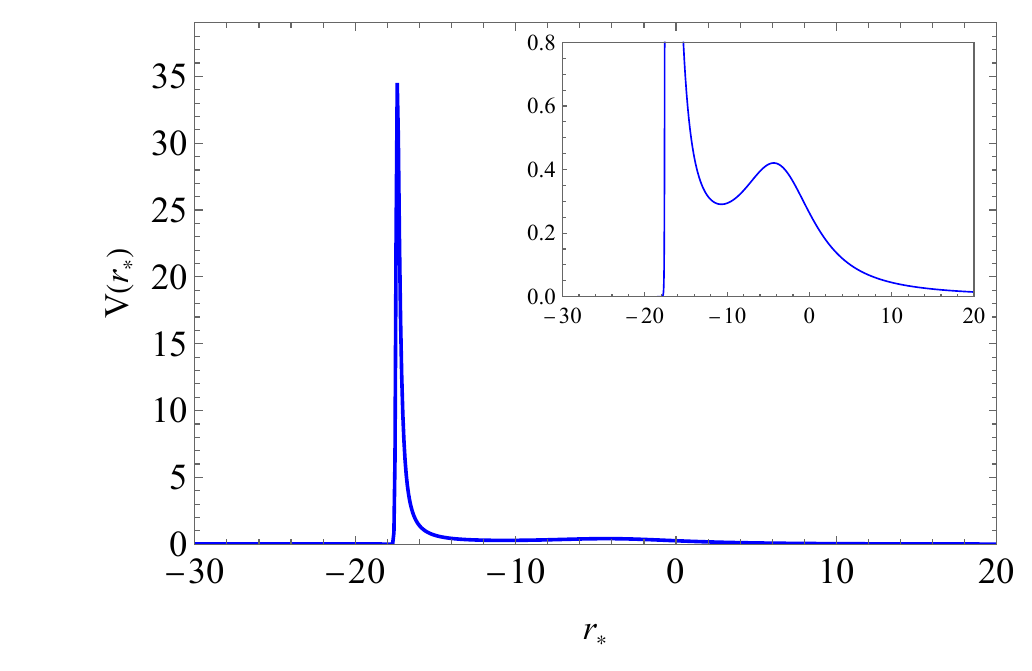}
\hspace{-1.1cm}
\includegraphics[width=0.55\textwidth]{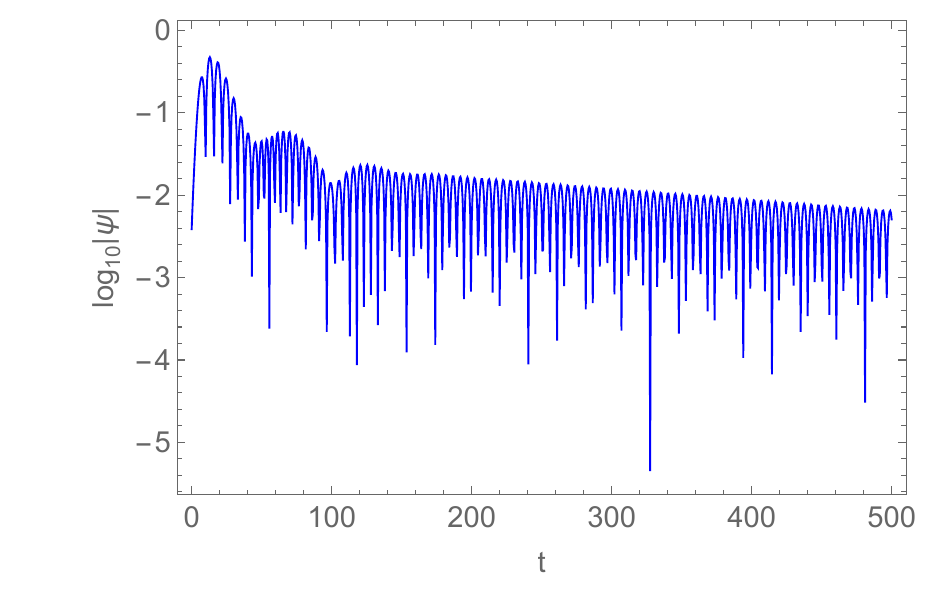}}
\makebox[\textwidth][c]{
\hspace{-1.0cm}
\includegraphics[width=0.55\textwidth]{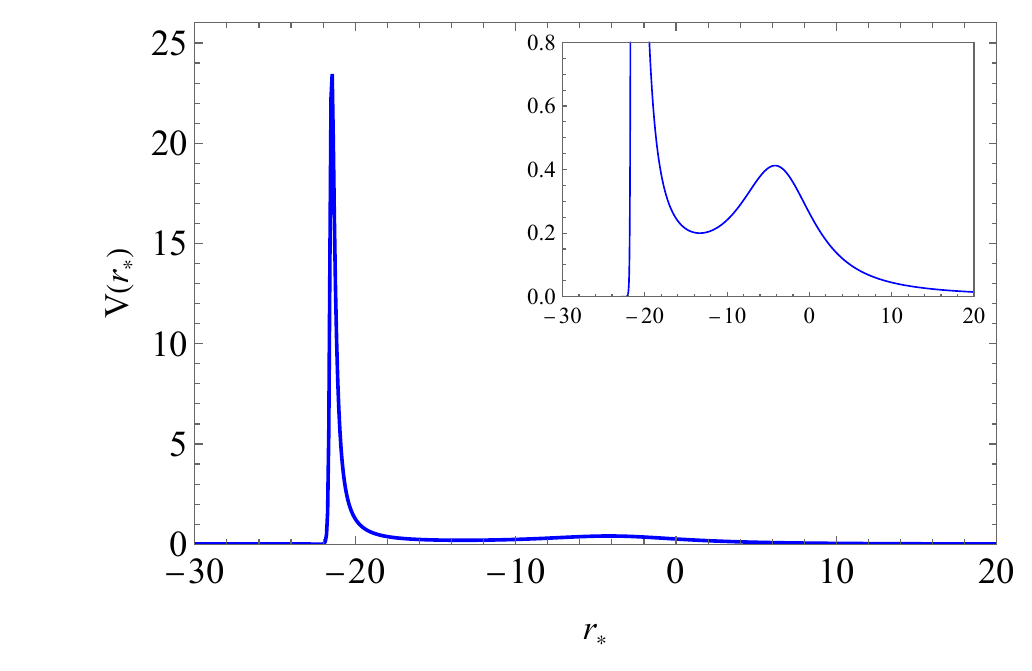}
\hspace{-1.1cm}
\includegraphics[width=0.55\textwidth]{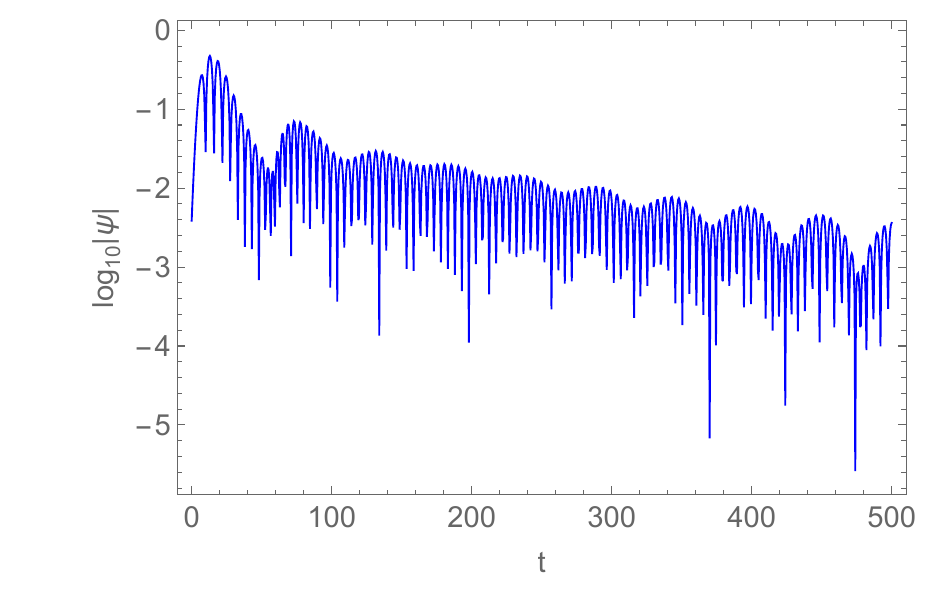}}
\makebox[\textwidth][c]{
\hspace{-1.0cm}
\includegraphics[width=0.55\textwidth]{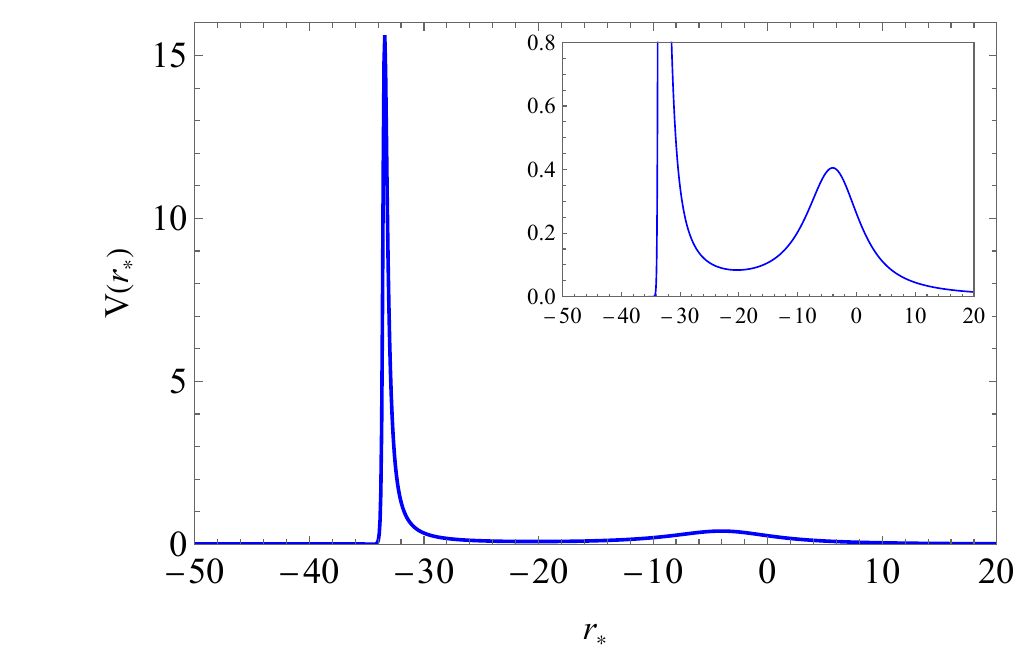}
\hspace{-1.1cm}
\includegraphics[width=0.55\textwidth]{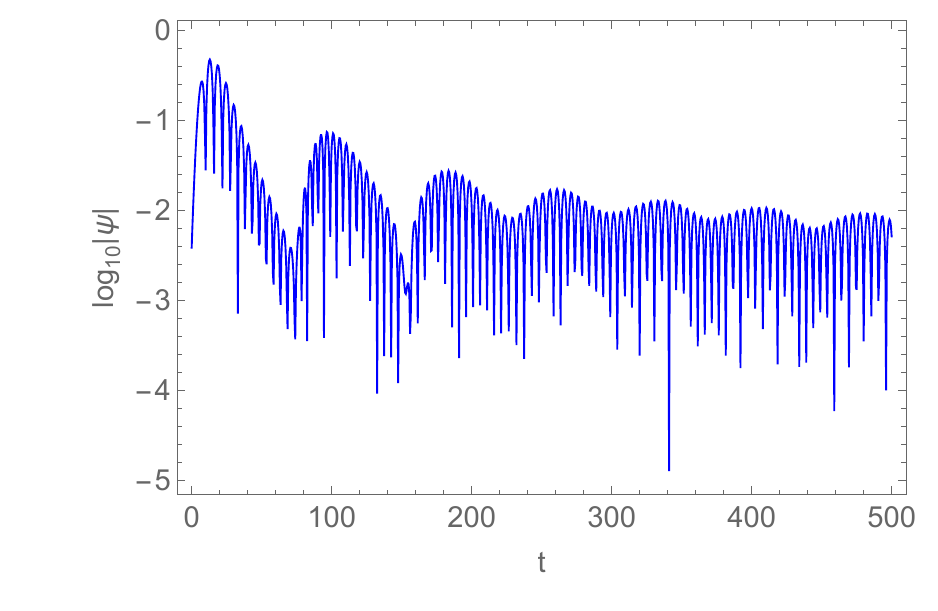}}
\end{center}
\setlength{\abovecaptionskip}{-0.2cm}
\setlength{\belowcaptionskip}{0.8cm}
\caption{Effective potential as a function of the tortoise coordinate $r_*$ (left panel), and the corresponding time-domain waveform as a function of time $t$ (right panel). The parameters are fixed at $M=1$, $\ell=2$, and $\alpha=0.45$. From top to bottom, the three curves correspond to $\beta=0.241$, $0.245$, and $0.249$, respectively.
}
\label{echo-alpha045}
\end{figure*}

Fig.~\ref{echo-beta025} shows the case of fixed $\beta=0.25$ with increasing $\alpha$. 
From the upper panel to the lower panel, $\alpha$ increases from $0.45$ to $0.46$. 
As $\alpha$ increases, the left peak of the effective potential becomes higher and shifts to larger $r_*$. 
Meanwhile, the distance between the two peaks decreases, and the cavity between the two barriers becomes narrower. 
The height of the right peak also decreases progressively. 
This trend is directly reflected in the time-domain profiles. 
For $\alpha=0.45$, the late-time waveform exhibits a relatively clear echo signal, and the interval between neighboring pulses is easy to identify. 
For $\alpha=0.455$, the delay becomes shorter, and the overlap between successive echoes becomes stronger. 
For $\alpha=0.46$, the signal becomes more compact, and the late-time waveform more closely resembles a strongly modulated long-lived ringing. 
Thus, in Fig.~\ref{echo-beta025}, increasing $\alpha$ reduces the peak value on the right side of the effective potential and causes it to gradually disappear, thereby resulting in the gradual weakening of the echo signal.
\begin{figure*}[htbp]
\begin{center}
\makebox[\textwidth][c]{
\hspace{-1.0cm}
\includegraphics[width=0.55\textwidth]{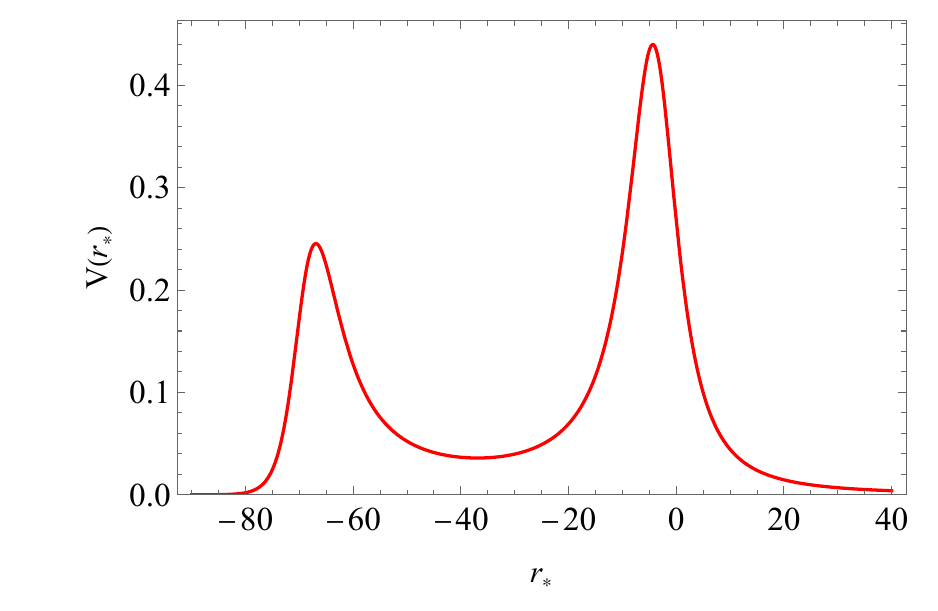}
\hspace{-1.1cm}
\includegraphics[width=0.55\textwidth]{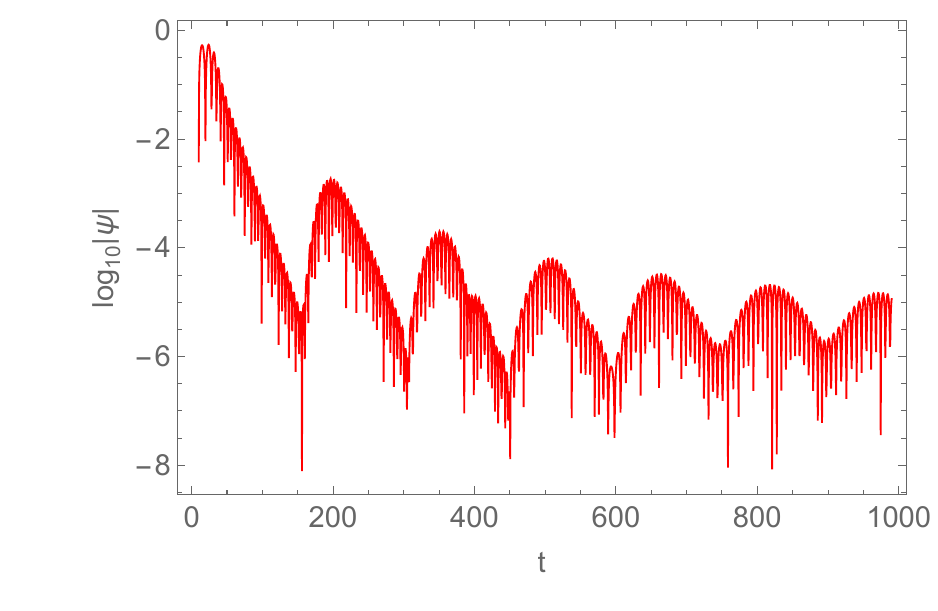}}
\makebox[\textwidth][c]{
\hspace{-1.0cm}
\includegraphics[width=0.55\textwidth]{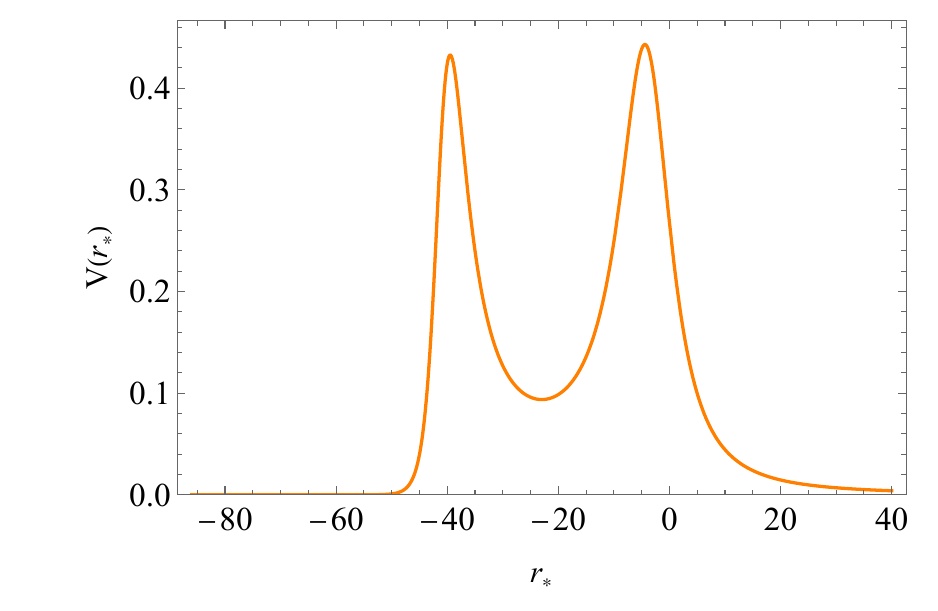}
\hspace{-1.1cm}
\includegraphics[width=0.55\textwidth]{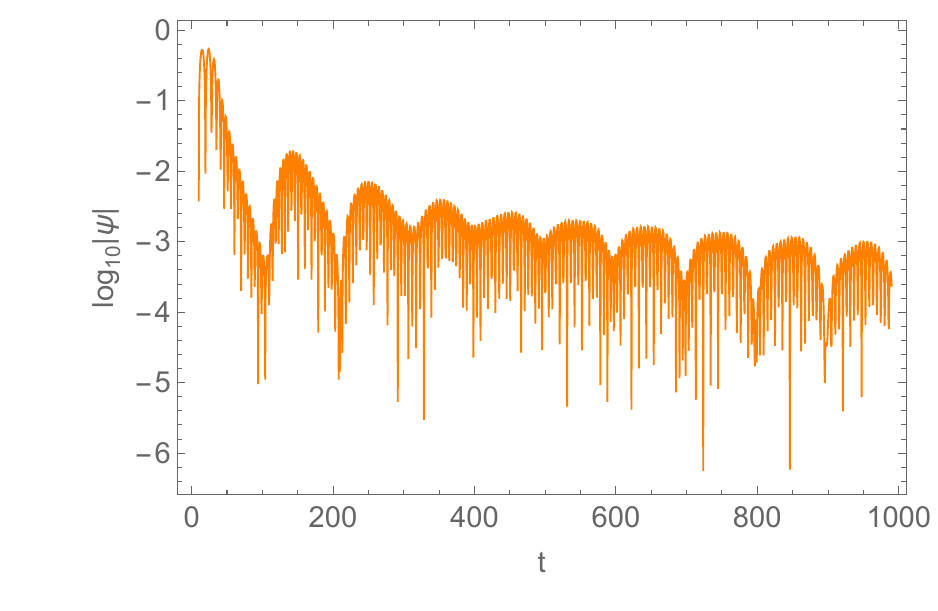}}
\makebox[\textwidth][c]{
\hspace{-1.0cm}
\includegraphics[width=0.55\textwidth]{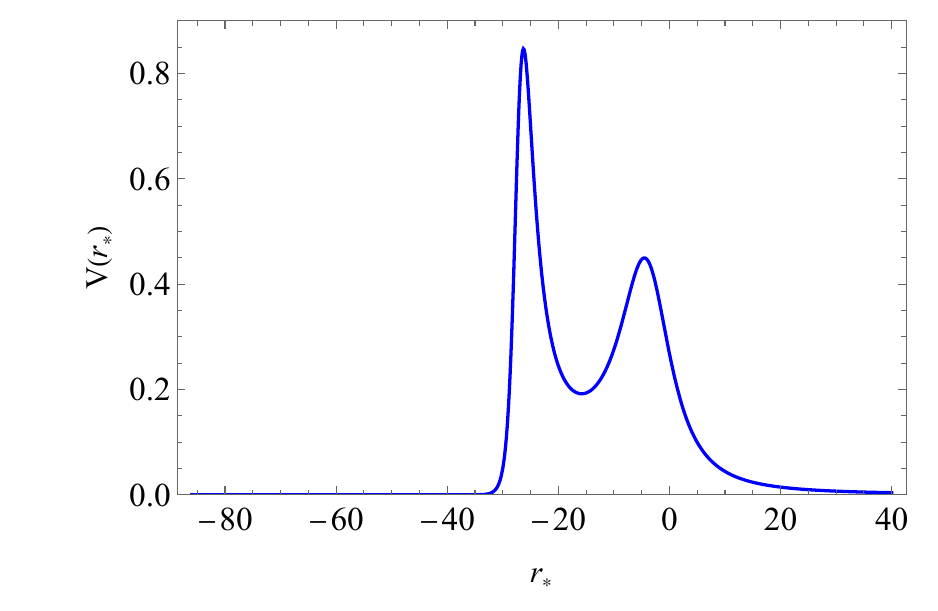}
\hspace{-1.1cm}
\includegraphics[width=0.55\textwidth]{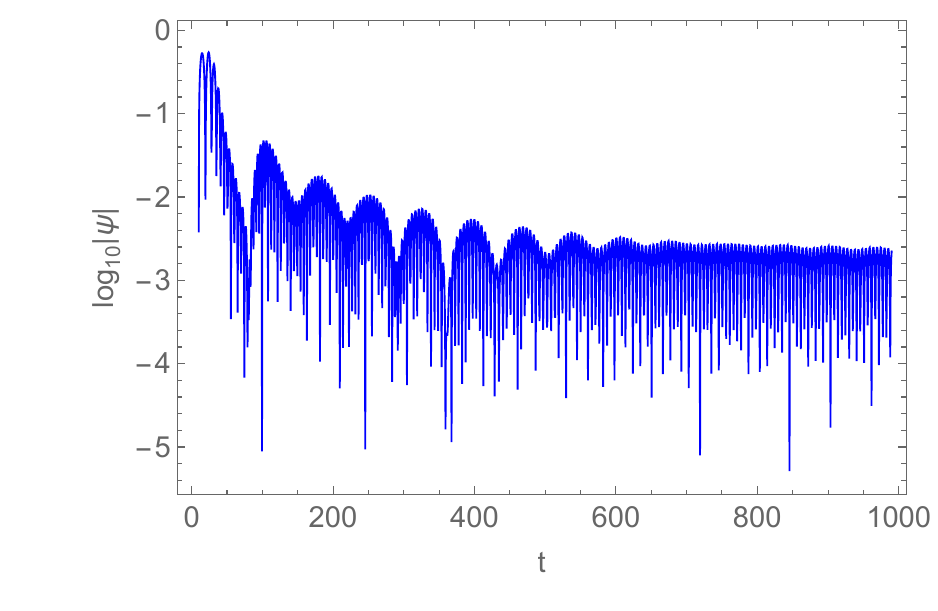}}
\end{center}
\setlength{\abovecaptionskip}{-0.2cm}
\setlength{\belowcaptionskip}{0.8cm}
\caption{Effective potential as a function of the tortoise coordinate $r_*$ (left panel) and the corresponding time-domain waveform (right panel). The parameters are fixed at $M=1$, $\ell=2$, and $\beta=0.296$. From top to bottom, the three cases correspond to $\alpha=0.493$ (top), $0.494$ (middle), and $0.496$ (bottom), respectively.
}
\label{echo-beta296}
\end{figure*}

Fig.~\ref{echo-alpha045} corresponds to the case of fixed $\alpha=0.45$ and varying $\beta$. 
From top to bottom, $\beta$ increases from $0.241$ to $0.249$. 
In this case, the inner peak moves to smaller $r_*$ and its height decreases, while the outer peak remains at a similar level. 
As a result, the cavity becomes wider, producing a clear change in the time-domain signal. 
For $\beta=0.241$, the waveform is dominated by a long-lived oscillatory signal with only weak large-scale modulation. 
For $\beta=0.245$, a greater number of faint echo signals can be observed. 
For $\beta=0.249$, the echo pattern becomes much clearer, and the spacing between successive pulses becomes larger. 
Furthermore, it should be noted that the trends of the effective potential and the echo signal are opposite to those shown in Fig.~\ref{echo-beta025}.

In Fig.~\ref{echo-beta296}, we fix $\beta=0.296$ and vary $\alpha$. 
From top to bottom, $\alpha$ increases from $0.493$ to $0.496$. 
The left peak grows rapidly and shifts toward larger $r_*$. 
Meanwhile, the distance between the two peaks decreases. 
The potential cavity is still present, but it becomes shorter and more strongly confined. 
This change is again visible in the time-domain waveforms. 
For $\alpha=0.493$, the signal shows well-separated late-time pulses. 
For $\alpha=0.494$, the pulses become closer, and the modulation becomes denser. 
For $\alpha=0.496$, the echo interval becomes even shorter, and the late-time signal approaches a quasi-periodic trapped oscillation. 
Importantly, the behavior of the echoes in Fig.~\ref{echo-beta296} is qualitatively similar to that in Fig.~\ref{echo-beta025}, while the corresponding effective potentials exhibit different trends.

\begin{figure*}[htbp]
\begin{center}
\makebox[\textwidth][c]{
\hspace{-1.0cm}
\includegraphics[width=0.55\textwidth]{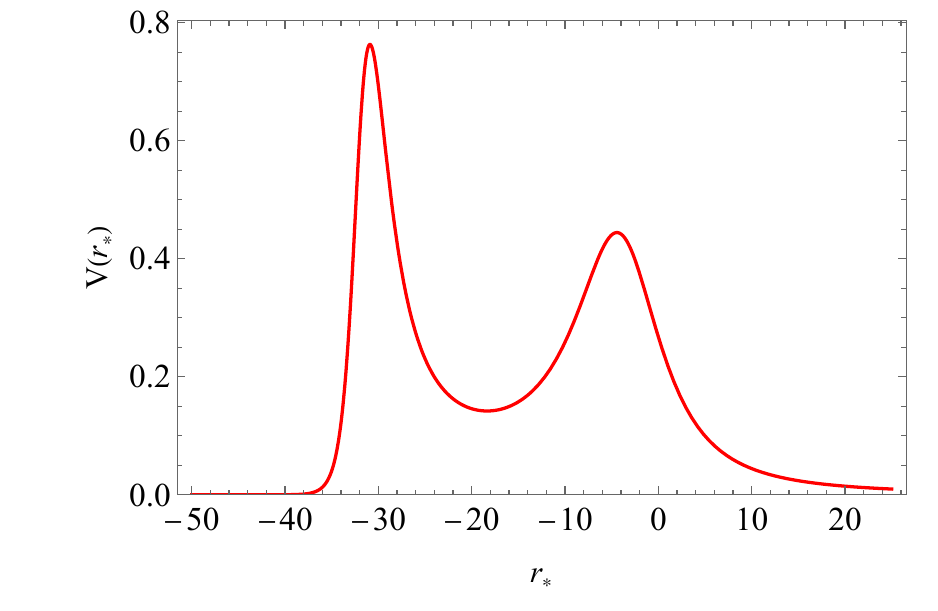}
\hspace{-1.1cm}
\includegraphics[width=0.55\textwidth]{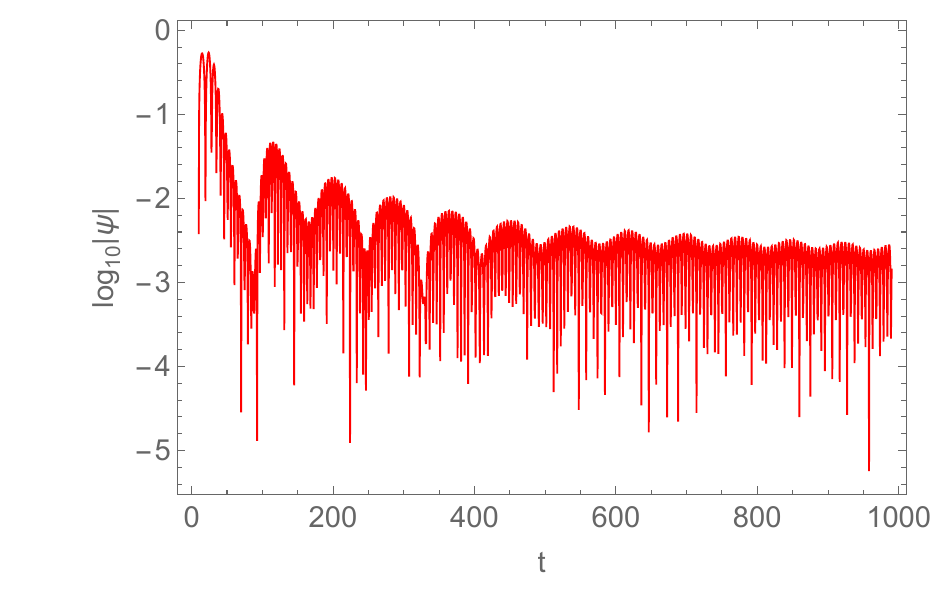}}
\makebox[\textwidth][c]{
\hspace{-1.0cm}
\includegraphics[width=0.55\textwidth]{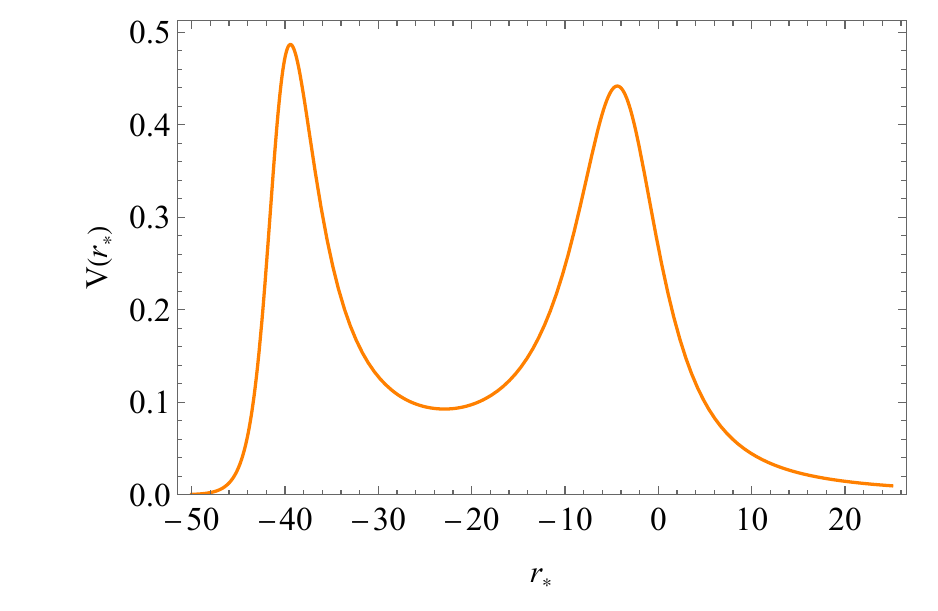}
\hspace{-1.1cm}
\includegraphics[width=0.55\textwidth]{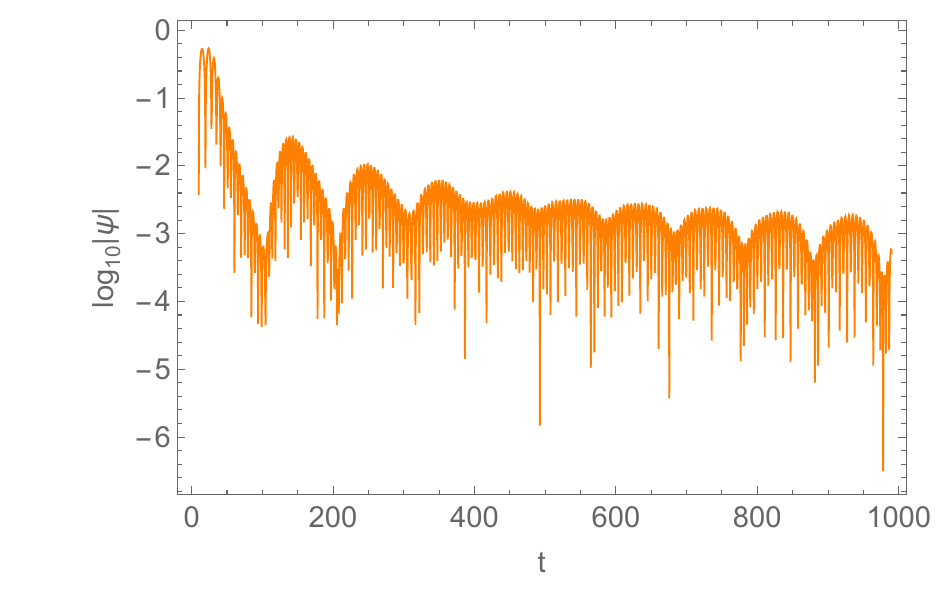}}
\makebox[\textwidth][c]{
\hspace{-1.0cm}
\includegraphics[width=0.55\textwidth]{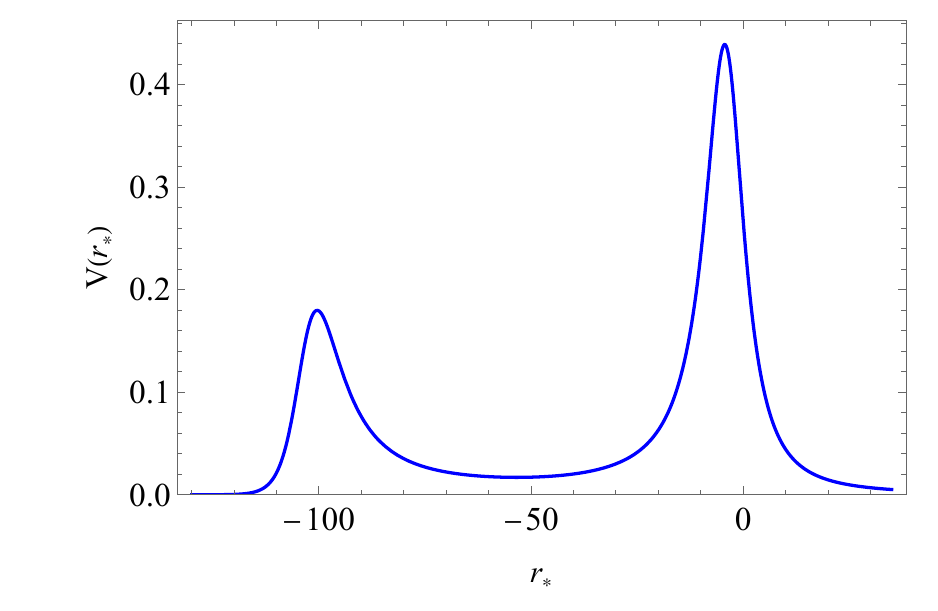}
\hspace{-1.1cm}
\includegraphics[width=0.55\textwidth]{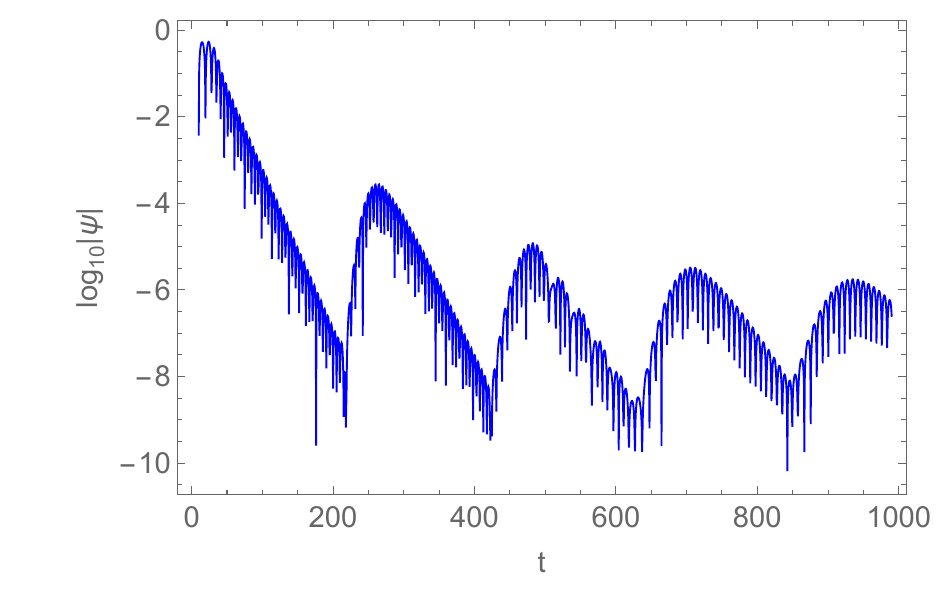}}
\end{center}
\setlength{\abovecaptionskip}{-0.2cm}
\setlength{\belowcaptionskip}{0.8cm}
\caption{Effective potential as a function of the tortoise coordinate $r_*$ (left panel) and the corresponding time-domain waveform (right panel). The parameters are fixed at $M=1$, $\ell=2$, and $\alpha=0.493$. From top to bottom, the three cases correspond to $\beta=0.294$, $0.295$, and $0.2963$, respectively.
}
\label{echo-alpha493}
\end{figure*}

Fig.~\ref{echo-alpha493} presents the case of fixed $\alpha=0.493$ and varying $\beta$. 
From top to bottom, $\beta$ changes from $0.294$ to $0.2963$. 
A clear rearrangement of the double-peak potential can be seen in the left panels. 
As $\beta$ increases, the left barrier decreases, while the right barrier increases. 
At the same time, the distance between the two peaks increases. 
Therefore, the effective cavity broadens. 
The time-domain signals show a similar trend. 
For $\beta=0.294$, the late-time waveform is still dominated by a relatively compact modulated oscillation. 
For $\beta=0.295$, the secondary pulses become easier to distinguish. 
For $\beta=0.2963$, the echo train is much more clearly resolved, and the delay between neighboring echoes is significantly larger. 
This figure therefore shows that, in the present case, an increase in the cavity width is accompanied by a more pronounced echo structure and a longer echo timescale.

The above results clearly show that the effective potential structure in the region above the shaded area and near the cusp point in Fig.~\ref{alpha_beta} is qualitatively different from that in the region farther from the cusp point. 
However, in both the fixed-$\alpha$ and fixed-$\beta$ cases, the echo behavior follows a similar pattern. 
This finding is important because it suggests that echo behavior is a robust feature of the hairy black hole spacetime in the double-peak regime and does not rely on a unique form of the effective potential.

To further understand the delayed structures shown in the time-domain profiles, we estimate the characteristic propagation time associated with the cavity formed by the two maxima of the double-peak effective potential.
The double-peak structure of the axial effective potential provides a natural geometric timescale for the delayed response of the waveform. When \(V_{\rm ax}(r)\) has two local maxima, denoted by \(r_{\rm L}\) and \(r_{\rm R}\) with \(r_{\rm L}<r_{\rm R}\), the region between them acts as an effective trapping cavity. A wave packet can be partially reflected back and forth inside this cavity before leaking out through the outer barrier, as in other effective-cavity descriptions of echo-like signals~\cite{Rosato:2025byu,Rosato:2025lxb}.
The relevant distance is not the coordinate distance in \(r\), but the distance measured in the tortoise coordinate. The tortoise distance between the two potential barriers is
\begin{equation}
\Delta r_*
=
\left|r_*(r_{\rm R})-r_*(r_{\rm L})\right|
=
\left|
\int_{r_{\rm L}}^{r_{\rm R}} e^{-\nu(r)}dr
\right| .
\end{equation}
Therefore, the round-trip propagation time inside the cavity can be estimated as
\begin{equation}
\Delta t_{\rm echo}^{\rm geo}
\simeq
2\Delta r_*
=
2\left|
\int_{r_{\rm L}}^{r_{\rm R}} e^{-\nu(r)}dr
\right| .
\label{echo_delay_geo}
\end{equation}
This quantity gives the characteristic timescale associated with the double-peak potential and repeated propagation within the effective cavity. This geometric estimate is expected to characterize the time delay between successive echoes when the two potential barriers are sufficiently well separated~\cite{Guo:2022umh,Sang:2022hng}. However, when the barrier spacing is small, the delayed wave packets may overlap significantly, making it difficult to define a clear interval between pulses; in such cases, the geometric approximation~(\ref{echo_delay_geo}) becomes less accurate~\cite{Guo:2022umh}.

\section{Trapping, light rings, and stability}
\label{light-ring}

The double-peak axial potential has a corresponding null-geodesic structure in the eikonal limit. For a metric of the form $ds^2=f(r)dt^2-f(r)^{-1}dr^2-r^2d\Omega^2$, equatorial null geodesics are governed, up to conserved energy and angular momentum factors, by
\begin{equation}
U_{\rm null}(r)=\frac{f(r)}{r^2}.
\end{equation}
Circular null orbits, or light rings, are extrema of $U_{\rm null}$ and therefore satisfy
\begin{equation}
r f'(r)-2f(r)=0.
\label{eq:light-ring-general}
\end{equation}
For the present logarithmic metric, this equation becomes
\begin{equation}
-2+\frac{6M}{r}
-\frac{\alpha M}{r^2}
\left[3+4\ln\!\left(\frac{r}{\beta}\right)\right]=0.
\label{eq:light-ring-log}
\end{equation}
A circular null orbit is radially stable when $U_{\rm null}''(r)>0$ and unstable when $U_{\rm null}''(r)<0$. Moreover, the axial potential satisfies
\begin{equation}
V_{\rm ax}(r)=\ell(\ell+1)U_{\rm null}(r)+\mathcal{O}(\ell^0),
\qquad \ell\rightarrow\infty,
\label{eq:eikonal-light-ring}
\end{equation}
so that, in the eikonal limit, two potential maxima separated by a minimum correspond to two unstable light rings separated by a stable one. At finite multipole number, and in particular for $\ell=2$, the extrema of $V_{\rm ax}$ are shifted relative to the null-geodesic radii by the subleading terms in Eq.~\eqref{axial_potential}.

A direct numerical check shows that all representative echo configurations displayed in Figs.~\ref{echo-beta025}--\ref{echo-alpha493} exhibit this exterior unstable--stable--unstable light-ring sequence. For example, for $M=1$, $\alpha=0.45$, and $\beta=0.25$, the event horizon is located at $r_h\simeq0.21822$, while Eq.~\eqref{eq:light-ring-log} yields
\begin{equation}
r_{\rm LR}^{(1)}\simeq0.26217\;(\mathrm{unstable}),\qquad
r_{\rm LR}^{(2)}\simeq0.67358\;(\mathrm{stable}),\qquad
r_{\rm LR}^{(3)}\simeq1.43108\;(\mathrm{unstable}).
\end{equation}
For the corresponding $\ell=2$ axial potential, the two maxima and the intervening minimum occur at $r\simeq0.27517$, $0.67454$, and $1.51829$, respectively. The stable light ring is therefore closely associated with the trapping structure in this model: in the eikonal limit, it provides the null-geodesic counterpart of the trapping region and suggests the possibility of long-lived, quasi-bound cavity excitations. This interpretation is consistent with the photon-sphere and echo-family classification developed for a different double-peak hairy-black-hole system in Ref.~\cite{Yang:2026qmf}.

The presence of a stable light ring or a potential well is, however, not in itself proof of either linear or nonlinear instability~\cite{Cunha:2017eoe,Cunha:2022gde}. With our convention $e^{-i\omega t}$, a linearly growing mode would satisfy $\operatorname{Im}(\omega)>0$. The single-barrier modes reported in Table~\ref{qnf-alpha} have $\operatorname{Im}(\omega)<0$, and the double-barrier time evolutions shown above exhibit leakage and decay over the simulated interval. These observations provide no evidence for a rapidly growing axial mode, but they do not constitute a proof of global stability. In particular, we have not computed the complete QNM spectrum or the eigenfunction localization in the double-peak branch; thus, a very weakly growing mode, a mode whose excitation depends strongly on the initial data or observation window, or a nonlinear secular effect cannot be excluded on the basis of the present finite-time calculation. A definitive stability analysis would require a full frequency-domain search for all double-barrier modes, together with perturbations of the effective source, the polar sector, and ultimately the nonlinear backreaction. We therefore make the limited claim that the geometry supports axial trapping and echo-like signals, without claiming complete linear or nonlinear stability of the underlying hairy solution.

\section{Conclusion}
\label{sec:conclusion}
In this work, we studied axial gravitational perturbations of a hairy black hole generated within the gravitational-decoupling framework. We reviewed the background geometry, its horizon structure, and the role of the hair parameters $\alpha$ and $\beta$. We then derived the odd-parity master equation and obtained the corresponding effective potential for axial gravitational perturbations.

In the single-barrier regime, we computed the fundamental quasinormal frequencies using three complementary methods: the pseudospectral approach, the higher-order WKB approximation, and Prony extraction from the time-domain signal. The results are in good overall agreement and show that the real part of the frequency increases with the hair parameter, while the damping rate exhibits a nonmonotonic dependence on $\alpha$.

Our main result is that, within a suitable region of parameter space, the axial effective potential develops a double-peak structure. This structure creates a trapping cavity and gives rise to echo-like late-time signals in the time-domain waveform. In this picture, the delayed pulses are not produced by an ad hoc reflecting surface near the horizon, but arise dynamically from the geometry of the effective potential itself. The echo timescale and the degree of pulse separation are controlled by the relative height and separation of the two barriers.

Overall, the present analysis shows that gravitational decoupling can produce hairy black hole backgrounds whose axial sector displays both modified quasinormal ringing and echo-like late-time behavior. The null-geodesic analysis further shows that the representative double-peak configurations contain a stable light ring between two unstable light rings. In the eikonal limit, this structure provides the geometric counterpart of the axial trapping cavity and can support long-lived cavity excitations. Nevertheless, the present study is linear, restricted to the fixed-source axial sector, and does not determine the complete double-barrier spectrum, the polar or coupled matter perturbations, or the nonlinear backreaction. The decaying waveforms found here therefore demonstrate an echo-supporting axial cavity, but they should not be interpreted as proof of complete stability. Extending the analysis to the full mode-family structure, coupled perturbations, and nonlinear evolution is an important direction for future work.

\begin{acknowledgments}
This research was funded by the Guizhou Provincial Basic Research Program (Natural Science) Youth Guidance Program  (No.~QN [2025] 365), the Guizhou Provincial Basic Research Program General Project (No.~MS [2026] 068), the National Natural Science Foundation of China (No.~12505064), the project of Young Scientific and Technical Talents Development of Education Department of Guizhou Province (No.~[2024] 79), and the Guizhou Provincial Basic Research Program (Natural Science) under Grant No.~QN [2025] 310. A. \"O. would like to acknowledge networking support of the COST Action CA21106 - COSMIC WISPers in the Dark Universe: Theory, astrophysics and experiments (CosmicWISPers), the COST Action CA22113 - Fundamental challenges in theoretical physics (THEORY-CHALLENGES), the COST Action CA21136 - Addressing observational tensions in cosmology with systematics and fundamental physics (CosmoVerse), the COST Action CA23130 - Bridging high and low energies in search of quantum gravity (BridgeQG), and the COST Action CA23115 - Relativistic Quantum Information (RQI) funded by COST (European Cooperation in Science and Technology). A. \"O. also thanks EMU, TUBITAK, ULAKBIM (Turkiye) and SCOAP3 (Switzerland) for their support.

\end{acknowledgments}

\bibliography{ref}
\bibliographystyle{apsrev4-1}

\end{document}